\documentclass[conference]{IEEEtran}
\IEEEoverridecommandlockouts
\usepackage{cite}
\usepackage{amsmath,amssymb,amsfonts}
\usepackage{algorithmic}
\usepackage{graphicx}
\usepackage{textcomp}

\usepackage{xcolor}
\usepackage[labelsep=period]{caption}

\usepackage{subcaption}
\usepackage{url}
\def\BibTeX{{\rm B\kern-.05em{\sc i\kern-.025em b}\kern-.08em
    T\kern-.1667em\lower.7ex\hbox{E}\kern-.125emX}}

\DeclareCaptionFont{xipt}{\fontsize{8} {10}\mdseries}
\usepackage[font=xipt]{caption}
\usepackage[font={small,rm}]{subcaption}

\graphicspath{{./}{Figures/}}

\newcommand{\myfigwidth}{5.5cm}

\begin{document}

\title{Rendering Spatial Sound for Interoperable Experiences in the Audio Metaverse}

\author{
\IEEEauthorblockN{Jean-Marc Jot}
\IEEEauthorblockA{\textit{iZotope, Inc.} \\
Aptos, CA USA \\
jjot@izotope.com}
\and
\IEEEauthorblockN{Rémi Audfray}
\IEEEauthorblockA{\textit{Facebook Reality Labs} \\
San Francisco, CA USA \\
remiaudfray@fb.com}
\and
\IEEEauthorblockN{Mark Hertensteiner}
\IEEEauthorblockA{\textit{Magic Leap, Inc.} \\
San Jose, CA USA \\
mhertensteiner@magicleap.com}
\and
\IEEEauthorblockN{Brian Schmidt}
\IEEEauthorblockA{\textit{Brian Schmidt Studios, LLC.} \\
Bellevue, WA USA \\
brian@brianschmidtstudios.com}
}

\maketitle

\begin{abstract}
Interactive audio spatialization technology previously developed for video game authoring and rendering has evolved into an essential component of platforms enabling shared immersive virtual experiences for future co-presence, remote collaboration and entertainment applications. New wearable virtual and augmented reality displays employ real-time binaural audio computing engines rendering multiple digital objects and supporting the free navigation of networked participants or their avatars through a juxtaposition of environments, real and virtual, often referred to as the Metaverse. These applications require a parametric audio scene programming interface to facilitate the creation and deployment of shared, dynamic and realistic virtual 3D worlds on mobile computing platforms and remote servers.

We propose a practical approach for designing parametric 6-degree-of-freedom object-based interactive audio engines to deliver the perceptually relevant binaural cues necessary for audio/visual and virtual/real congruence in Metaverse experiences. We address the effects of room reverberation, acoustic reflectors, and obstacles in both the virtual and real environments, and discuss how such effects may be driven by combinations of pre-computed and real-time acoustic propagation solvers. We envision an open scene description model distilled to facilitate the development of interoperable applications distributed across multiple platforms, where each audio object represents, to the user, a natural sound source having controllable distance, size, orientation, and acoustic radiation properties.
\footnote{Originally presented at the International Conference on Immersive and 3D Audio (Sep 2021). Rémi Audfray was with Magic Leap when this paper was authored and submitted for publication.}
\end{abstract}


\section{From 3DoF to 6DoF Object-Based Audio}

\subsection{Recent Developments in 3-Degree-of-Freedom Media}
\label{sec:media}
Spatial audio reproduction technology has developed since the middle of the 20th century from \textit{mono} to \textit{stereo} to \textit{surround} to \textit{immersive} \cite{roginska2017}. George Lucas once said that audio accounts for “half the experience” when it is implemented in conjunction with visual reproduction in the cinema \cite{blake2004}. More recently, thanks to the adoption of new immersive audio formats in the movie industry, it has become possible to achieve a greater degree of spatial coincidence between the on-screen visual cues and the audio localization cues delivered via loudspeakers to a fixed (seated) spectator in a theater.

These new multi-channel audio formats are also employed in the creation of cinematic Virtual Reality (VR) experiences and their delivery via Head-Mounted Display devices (HMD) which incorporate head-tracking technology enabling 3-degree-of-freedom (3DoF) experiences in which movements in the spectator's head orientation are compensated for in both the visual and audio presentations. These new forms of linear media have begun introducing mainstream consumers and the entertainment industry to immersive and positional audio concepts, and to new capabilities over previous stereo or multi-channel audio formats \cite{robinson2012atmos, herre2014mpegh, adm}:
\begin{itemize}
    \item extended positional audio playback resolution and coverage (including above or below the horizontal plane), as well as spatially diffuse reverberation rendering
    \item flexible loudspeaker playback configurations, and support for encoding and rendering an audio scene in High-Order Ambisonic format (HOA) \cite{zotter2019}
    \item ability to substitute or modify selected elements of the audio mix (encoded as \textit{audio objects}) for content customization such as language localization or hearing enhancement \cite{shirley2015}.
\end{itemize}

\begin{figure} [b] 
    \centering
    \includegraphics[width=0.65\columnwidth]{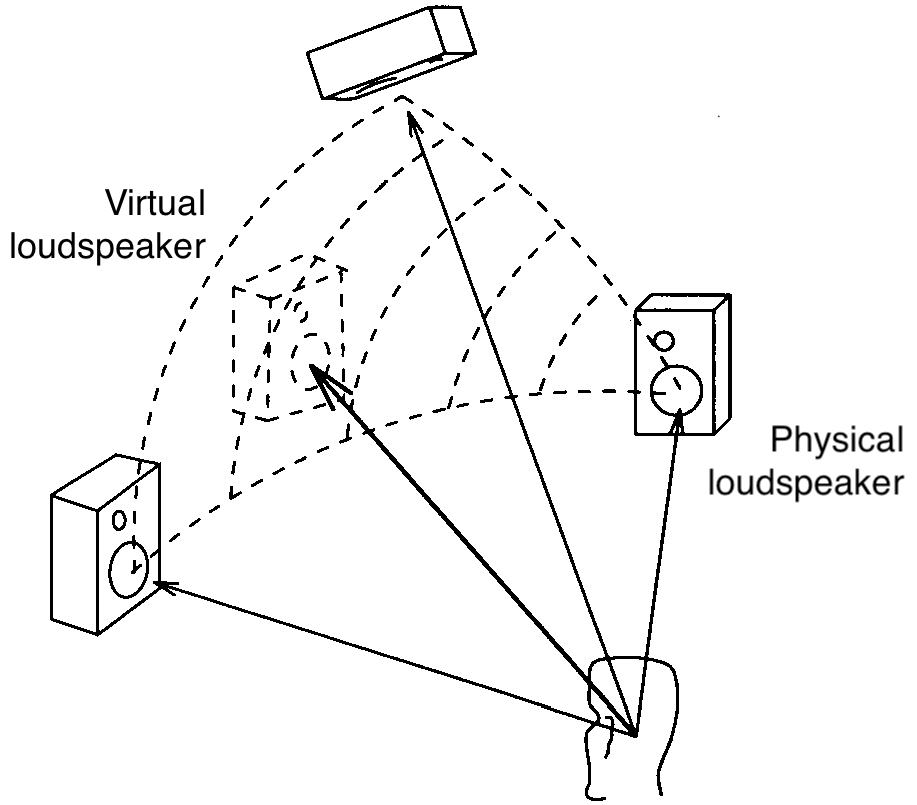}
    \caption{In 3DoF object-based immersive audio formats, each mono-waveform object represents a virtual loudspeaker assigned a variable angular position \cite{robinson2012atmos, herre2014mpegh, adm} (drawing adapted from \cite{pulkki1997}).}
    \label{fig:3dofScene}
\end{figure}

Some limitations persist.  For instance, it is not practical, in current 3DoF immersive audio formats, to substitute objects that carry spatially diffuse reverberation or to allow free listener navigation "inside the audio mix” at playback time. Essentially, a 3DoF audio object represents a virtual loudspeaker oriented towards the listener and allowed variable position on a surrounding surface, as illustrated in Fig. \ref{fig:3dofScene}.

\newpage
\subsection{Towards the 6-Degree-of-Freedom Audio Metaverse}

In interactive Virtual Reality or Augmented Reality (AR) experiences, the audio-visual presentation must also account for possible translations of the user’s position along up to three coordinate axes $(x,y,z)$, simulating 6-Degree-of-Freedom (6DoF) navigation inside a virtual world, hereafter referred to as the \textit{Metaverse} \cite{stephenson1992snowcrash, ball2021}.
Like the physical world, the Metaverse may be modeled, from a human perception and cognition standpoint, as a collection of audio-visual objects (each a sound source and/or a light emitter or reflector) coexisting in an environment that manifests itself to the spectator through ambient lighting and acoustic reflections or reverberation.

Unlike 3DoF experiences, 6DoF scenarios may allow a user to walk closer to an object or around it, or may allow an object to move freely in position or orientation (in other words, audio objects may also be allowed 6 degrees of freedom). The acoustic properties of an audio object should include its directivity pattern, so that sound intensity and spectral changes may be simulated convincingly upon variation in its orientation relative to the position of the listener. Existing 6DoF scene representation formats and technologies for authoring and rendering navigable virtual worlds include VRML (for web applications), MPEG-4 (for synchronized multimedia delivery), and game engines \cite{vrml, scheirer1999, game_audio}. New standards under development -- such as OpenXR, WebXR, and MPEG-I -- address modern VR and AR content and devices, including support for listener navigation between multiple Ambisonic scene capture locations \cite{openxr, webxr, quackenbush2021, eduardo2019}.

\subsection{The Coming-of-Age of Wearable 3D Audio Appliances}

On the device side, several recent technology innovations accelerate the deployment of 6DoF audio technology \cite{rumsey2019headphone}:
\begin{itemize}
    \item New 3DoF listening experiences and devices have begun introducing mainstream consumers to head-tracking technology (as an enhancement beyond traditional "head-locked" headphone listening experiences), with or without coincident visual playback.
    \item The recent introduction of low-latency wireless audio transmission technology enables “unleashed” mobile listening experiences.
    \item Acoustically transparent binaural audio playback devices enable AR audio reproduction \cite{harma2004} -- including hear-through earbuds, bone conduction headsets, audio-only glasses, and audio-visual HMDs. 
\end{itemize}

In traditional headphone audio reproduction, binaural processing is often considered to be an optional function, aspiring to “help the listener forget that they are wearing headphones” by delivering a natural, externalized listening experience \cite{best2020}. Breaking from this legacy, the new generation of non-intrusive wireless wearable playback devices \textit{requires} binaural 3D audio processing technology, as a necessity to ensure that the audio playback system remains inconspicuous \cite{harma2004}.

\subsection{Shared Immersive Experiences in the Audio Metaverse} \label{sec:subsec6dofaudio}

Categories of applications facilitated by this technology evolution include:
\begin{itemize}
    \item Social co-presence, including virtual meetings, remote collaboration, and virtual assistants 
    \item Assisted training, education, diagnosis, surgery, design, or architecture
    \item Virtual travel, including virtual navigation or tourism, and situational awareness 
    \item Immersive entertainment, including games, music, and virtual concerts \cite{lunny2020}.
\end{itemize}

\begin{figure} [b]
    \centering
    \includegraphics[width=\columnwidth]{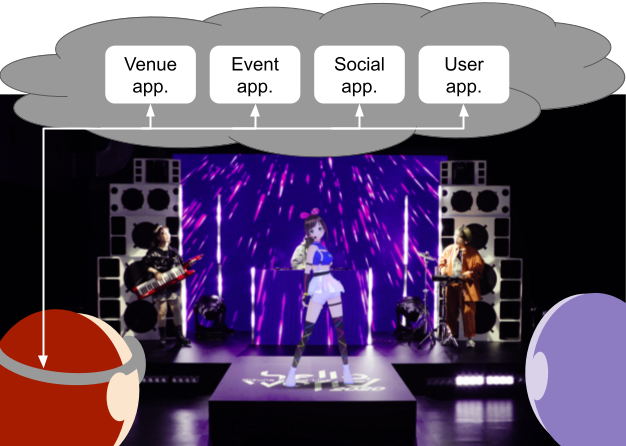}
    \caption{Example of 6DoF Metaverse experience: attending a virtual live music event including real and/or virtual audience members, performers and environmental elements, using a wearable AR display device connected to several cloud-hosted applications (image credit: restofworld.org/2021/vtubers).}
    \label{fig:6dofScene}
\end{figure}

In some of these use cases, as illustrated in Fig. \ref{fig:6dofScene}, a Metaverse portion or layer may be experienced simultaneously by multiple connected users navigating independently or actively engaged in communication or collaboration. Additionally, portions of the object or environment information may be served or updated at runtime by several independent applications employing a common syntax and protocol to transmit partial scene description data to be parsed, consolidated and processed by each local rendering device.

A general, versatile audio technology solution supporting these applications should satisfy both \textit{“you are there”} experiences (VR, teleportation to a virtual world) and \textit{“they are here”} use cases (AR, digitally extending the user’s physical world without obstructing their environment).  From an ecological point of view, an aim of the technology is to \textit{suspend disbelief}: alleviate the user’s cognitive effort by matching implicit expectations derived from natural experience. This translates, in particular, to the requirement of maintaining consistency between the audio and visual components of the scene experienced by the user (\textit{audio/visual congruence}) and between its virtual elements and the real world as experienced previously or concurrently (\textit{virtual/real congruence}).

\newpage
\subsection{Environmental Audio for VR and AR}

With regards to the modeling of acoustic reverberation and sound propagation, an application programming solution for the Audio Metaverse should meet the following criteria:
\begin{itemize}
    \item In VR applications, environment reverberation properties should be tunable by a sound designer via perceptually meaningful parameters, thereby extending the creative palette beyond the constraints and limitations inherent in physics-based environment simulation.
    \item In AR applications, the virtual acoustic scene representation should be adaptable to match or incorporate local environment parameters detected at playback time, including room reverberation attributes as well as the geometrical configuration and physical properties of real or virtual acoustic obstacles or reflectors.
\end{itemize}

Accordingly, the following capabilities should be enabled by a versatile programming and rendering solution:
\begin{itemize}
    \item \textit{At creation}: specify a collection of virtual sound sources and the intrinsic properties of each, including its acoustic directivity and a sound generation method (e.g. voice communication, procedural sound synthesis, etc. \cite{game_audio})
    \item \textit{At runtime}: according to application scenario and events, set source and listener positions and orientations, reverberation properties and acoustic simulation parameters to match local or virtual environment conditions.
\end{itemize}

\medskip
\subsection{Spatial Audio Rendering Technology for the Metaverse}

Binaural 3D audio rendering technology is essential in 6DoF virtual experiences mediated by wearable or head-mounted displays. In this paper, the requirements to be met by a successful solution will be broken down as follows:
\begin{itemize}
    \item \textit{Binaural rendering}: perceptually grounded audio signal processing engine designed to allocate rendering computation resources towards the minimum viable footprint, scalable to wearable or battery-powered devices (see methods reviewed in Section \ref{sec:rendering_algorithms} of this paper). Audio fidelity should be predictable and maximally consistent across platforms or conditions. The level-of-detail prioritization strategy should be based on perceptual significance ranking of the audio features.
    \item \smallskip \textit{Application development} facilitated by distinguishing two levels of Application Programming Interface (API):\\
    1) \textit{Low-level audio rendering API} reflecting the above psycho-acoustically based requirements. In Section \ref{sec:rendering_api}, we examine the progressive stages of elaboration of a practical and generic solution developed for this purpose, extending  PC and mobile game audio standards.\\
    2) \textit{High-level scene description API} distilled to the critical functional features required to optimize development efficiency. In Section \ref{sec:propagation}, we review practical acoustic propagation models and strategies which enable, from high-level scene description data, the derivation of the low-level audio rendering API parameters that control the digital audio scene delivery to each user.
\end{itemize}

\begin{figure} [t]
    \centering
    \includegraphics[width=\columnwidth]{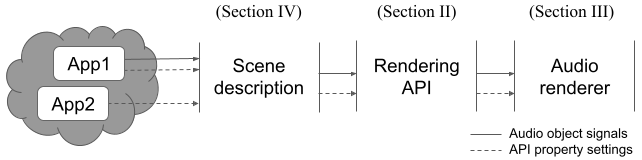} \caption{Audio rendering for the Metaverse: high-level functional system synopsis and organization of this paper.}
    \label{fig:paper_flow}
\end{figure}

Fig. \ref{fig:paper_flow} illustrates the proposed functional composition of the overall audio rendering system. By collecting this information in the present synopsis paper, we aim to promote cooperation towards the development of a practical multi-application, multi-platform Audio Metaverse data model coupled with an open acoustic rendering API, mature for industry-wide standardization to realize the promise of 6DoF spatial audio technology innovation outlined above in Section \ref{sec:subsec6dofaudio}.

\bigskip
\section{Audio Rendering API for the Metaverse} \label{sec:rendering_api}

Computer-generated audio reproduction technology for VR and AR leverages decades of signal processing innovation in interactive audio rendering systems and application programming interfaces -- building and extending upon prior developments in the fields of computer music and architectural acoustics (including binaural techniques, artificial reverberation algorithms, physical room acoustics modeling and auralization \cite{chowning1971, kleiner1993, jot1999acm, puckette2007}).

We begin this section with an overview of early developments in interactive audio standards for PC and mobile gaming, leading up to the OpenAL EFX API \cite{OpenALEFX}. We then highlight some of the new enhancements incorporated in Magic Leap's Soundfield Audio API (MSA), targeting today's AR and VR applications \cite{mlsdk, audfray2018audio}. Table \ref{table:APIcomparison} summarizes the evolution in successive capability extensions among these 6DoF object-based audio rendering APIs.

\begin{table*} [!htbp] 
\caption{Evolution of interactive object-based immersive audio rendering API capabilities.}
\centering
\small
\begin{tabular}{| l | c | c | c | c | c |}
    \hline \rule{0pt}{8pt}
    Audio rendering API features & (3DoF) & I3DL1, DSound & I3DL2, EAX2 & OpenAL EFX & MSA \\ 
    \hline\hline \rule{0pt}{8pt}
    Direction of arrival \textit{(per-object)} & $\bullet$ & $\bullet$ & $\bullet$ & $\bullet$ & $\bullet$ \\ 
    \hline \rule{0pt}{8pt} 
    Distance/proximity \textit{(per-object)} & $\circ$ & $\bullet$ & $\bullet$ & $\bullet$ & $\bullet$ \\ 
    \hline \rule{0pt}{8pt}
    Source orientation \textit{(per-object)} & $\circ$ & $\bullet$ & $\bullet$ & $\bullet$ & $\bullet$ \\ 
    \hline \rule{0pt}{8pt}
    Source directivity \textit{(per-object)} & $\circ$ & $\bullet$ & $\bullet$ & $\bullet$ & $\bullet$ \\ 
    \hline \rule{0pt}{8pt}
    Filtering effects \textit{(per-object)} & $\circ$ & $\circ$ & $\bullet$ & $\bullet$ & $\bullet$ \\ 
    \hline \rule{0pt}{8pt}
    Listener's room \textit{(global)} & $\circ$ & $\circ$ & $\bullet$ & $\bullet$ & $\bullet$ \\ 
    \hline \rule{0pt}{8pt}
    Multiple rooms \textit{(global)} & $\circ$ & $\circ$ & $\circ$ & $\bullet$ & $\bullet$ \\ 
    \hline \rule{0pt}{8pt}
    Clustered reflections \textit{(per-object)} & $\circ$ & $\circ$ & $\circ$ & $\circ$ & $\bullet$ \\ 
    \hline \rule{0pt}{8pt}
    Control frequencies \textit{(global)} & $\circ$ & $\circ$ & $Fh$ & $Fl$, $Fh$ & $Fl$, $Fm$, $Fh$ \\ 
    \hline
\end{tabular}
\label{table:APIcomparison}
\bigskip
\end{table*}

\bigskip
\subsection{Interactive Audio Standards for PC and Mobile Gaming}

Technology standards were created in the 1990s for PC audio, and later for mobile audio, to address the demand for interoperable hardware platforms and 3D gaming software applications, comparable in nature to the need that motivates present VR/AR industry initiatives such as OpenXR \cite{openxr}. Shared API standards were developed for graphics rendering (notably OpenGL) and, simultaneously, for audio rendering -- including I3DL2, EAX and OpenAL \cite{i3dl2, eax2, OpenAL}.

\medskip
\subsubsection{Positional audio properties} \hfill 

\begin{figure} [b] 
    \centering
    \includegraphics[width=\columnwidth]{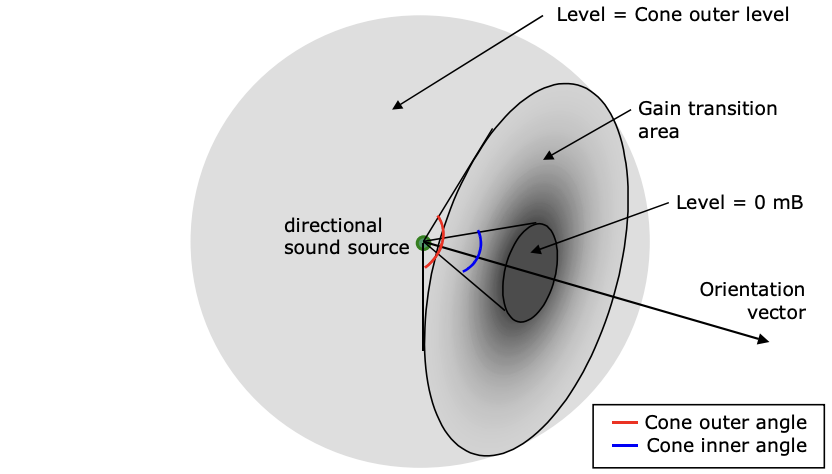}
    \smallskip
    \caption{Source directivity axis-symmetric "radiation cone" model (from \cite{Opensl}).}
    \label{fig:soundcone}
\end{figure}

In I3DL1/DirectSound3D, VRML97, OpenAL and OpenSL ES, the acoustic scene is described as a set of sound sources and a single listener, including \cite{vrml, i3dl1, OpenAL, Opensl}:
\begin{itemize}
    \item the \textit{position} and the \textit{orientation} of the listener and of each sound source, and a per-source gain correction
    \item \textit{velocity vectors} assigned to the listener and the sound sources, and a $Doppler\_factor$ property controlling Doppler effect strength (see e.g. \cite{OpenAL}, p. 28)
    \item automatic relative attenuation of sound as a function of source-listener distance beyond a $Reference\_distance$, customizable via a $Rolloff\_factor$ property
    \item a simplified \textit{source directivity} model -- see Fig. \ref{fig:soundcone} (note that more general radiation models were also proposed, e.g. in \cite{savioja1999, scheirer1999}).
\end{itemize}

\medskip
\subsubsection{Environmental audio extensions (single room)} \label{sec:reverbAPI} \hfill 

\begin{figure} [b]
    \centering
    \includegraphics[width=\columnwidth]{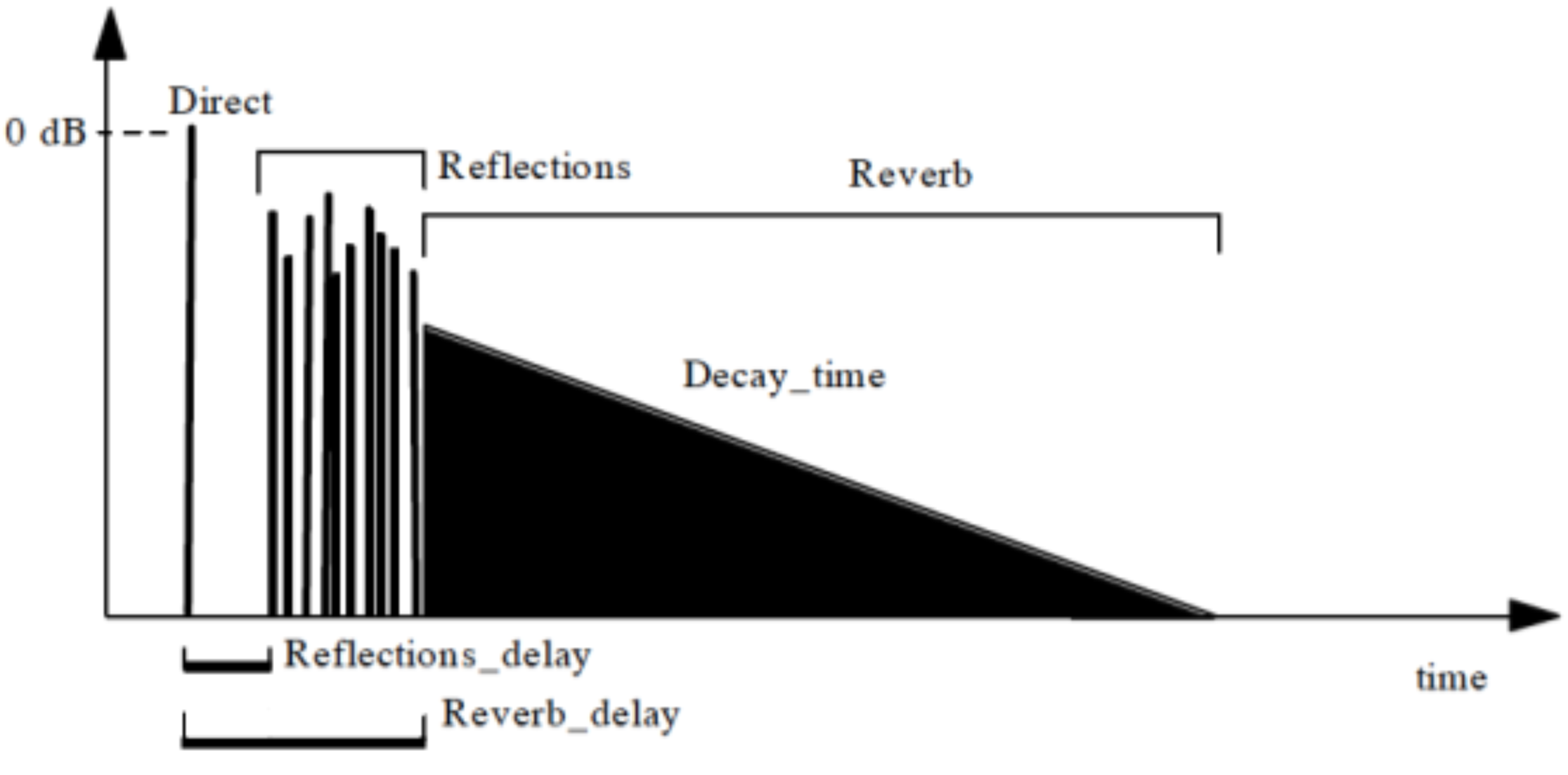}
    \caption{The Reverberation response model in Magic Leap Soundfield Audio (MSA) is compatible with I3DL2, EAX and OpenAL EFX \cite{mlsdk, i3dl2, OpenALEFX}.}
    \label{fig:RIR}
\end{figure}

I3DL2 and EAX 2.0 (1999), and then OpenAL EFX and OpenSL ES, included properties describing an "environment" (room) in which the listener and sound sources coexist. They decompose the rendered audio signal for each source into direct sound, early reflections and late reverberation, and expose the following features \cite{i3dl2, eax2, OpenALEFX, Opensl}:
\begin{itemize}
    \item room impulse response model (see Fig. \ref{fig:RIR}):
    \begin{itemize}
        \item a triplet of gain values ($Direct$, $Reflections$, $Reverb$), specified at the $Reference\_distance$
        \item onset times $Reflections\_delay$ and $Reverb\_delay$
        \item the reverberation's $Decay\_time$, $Density$, and $Diffusion$ (see Section \ref{sec:reverb_rendering})
    \end{itemize}
    \item reference frequency $Fh$ where high-frequency attenuations and the property $Decay\_hf\_ratio$ are specified
    \item per-source control of gain and high-frequency attenuation for $Direct$ and $Room$ (i.e. jointly Reflections and Reverb)
    \newpage
    \item spectral extension of the source radiation model: high-frequency attenuation $Outer\_gain\_hf$ controlling a low-pass filter effect dependent on source orientation (more details in Section \ref{sec:directivitymodel})
    \item globally defined air absorption model controlling a distance-dependent low-pass filter effect
    \item extension of the distance-based model to automatically apply a relative attenuation to the reflections and reverberation for distances beyond the $Reference\_distance$, including a per-source $Room\_rolloff\_factor$ property and a physical simulation mode (see Section \ref{sec:distancemodel})
    \item sound muffling effects for simulating sound transmission through a room partition ($Occlusion$) or through/around an obstacle inside the room ($Obstruction$).
\end{itemize}

\medskip
\subsubsection{Multi-room extensions} \hfill 

\begin{figure} [!htbp]
    \centering
    \includegraphics[width=\columnwidth]{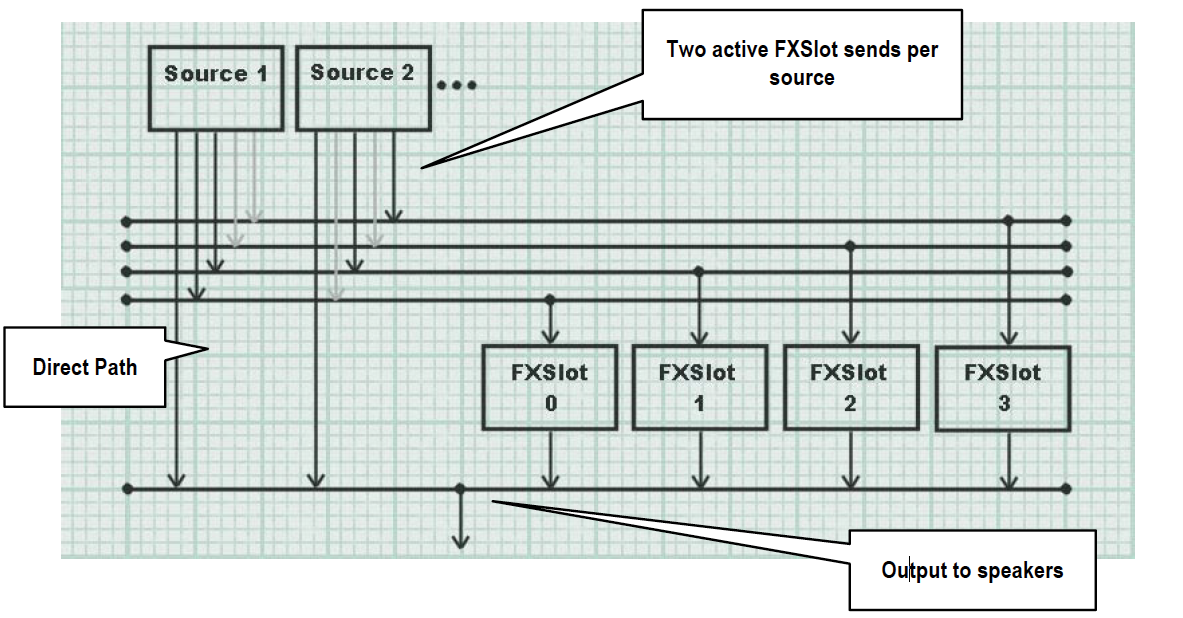}
    \caption{In EAX 4.0 and OpenAL EFX, each sound source can contribute into several auxiliary effect slots, each hosting a reverberator that simulates a room in the listener's surroundings and returns a multi-channel output signal into the main output bus (from \cite{eax4introduction}).}
    \label{fig:eax4signal}
\end{figure}

\begin{figure} [!htbp]
    \centering
    \medskip
    \includegraphics[width=0.6\columnwidth]{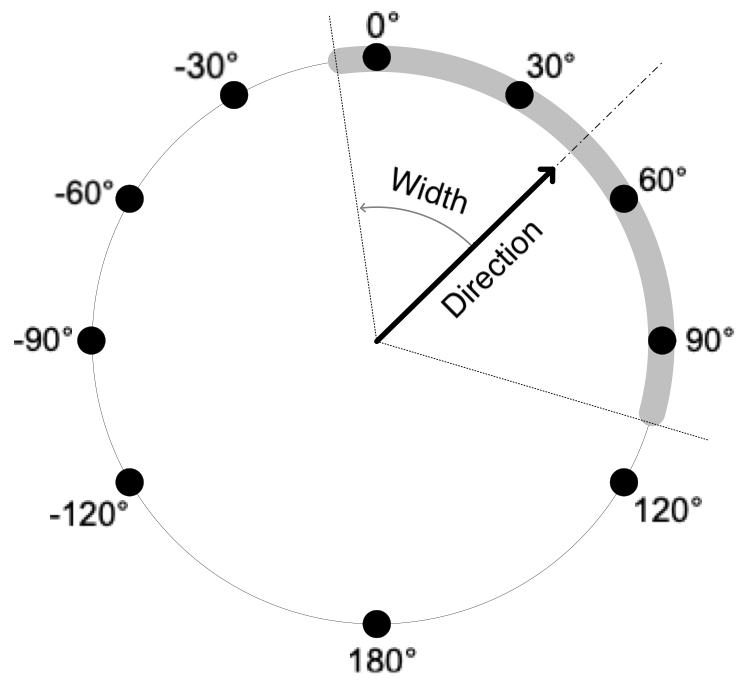}
    \caption{Illustration of the reverberation output panning model in EAX 4.0 and OpenAL EFX (from \cite{jot2006binaural}). The black circles represent head-locked virtual loudspeakers.}
    \label{fig:diffusepanning}
\end{figure}

EAX 4.0 and OpenAL EFX enabled loading up to four concurrent I3DL2/EAX reverberator plugins in "auxiliary effect slots," and exposed the following capabilities \cite{eax4programmer, OpenALEFX}:
\begin{itemize}
    \item selecting which of these four reverberators simulates the listener's environment at the current time
    \item separately controlled per-source contributions into several of these reverberators (see Fig. \ref{fig:eax4signal})
    \item reverberator output occlusion and diffuse panning model (illustrated in Fig. \ref{fig:diffusepanning})
    \item 3-band control of spectral effects using two reference frequencies, $Fl$ and $Fh$ (see Section \ref{sec:globalProps}).
\end{itemize}

\newpage
\subsection{From OpenAL to Magic Leap Soundfield Audio (MSA)}

OpenAL and its EFX extension cover a broad Section of the requirements of recent VR and AR applications for wearable interactive audio reproduction technology, and enable the binaural simulation of feature-rich audio scenes \cite{jot2006binaural}. In this section, we describe previously unpublished aspects of EAX/OpenAL functions, and how Magic Leap Soundfield Audio (MSA) expands upon these with the ambition to provide for the Audio Metaverse a complete yet efficient spatial rendering API -- including the following enhancements:
\begin{itemize}
    \item extended spectral control of source radiation pattern for more realistic modeling of 6DoF objects
    \item per-source spatialization of \textit{clustered reflections}, enabling improvements in the spatial imaging of distant audio objects and in geometry-driven acoustic simulation
    \item distance-dependent reflection and reverberation normalization and spectral attenuation model, exploiting the acoustical characterization of rooms by their \textit{reverberation fingerprint} \cite{jot2016augmented, audfray2019reverberation}.
\end{itemize}

\newpage
MSA also differs from OpenAL EFX by more explicitly distinguishing low-level vs. high-level API: listener geometry (position and orientation tracking) is handled in the high-level scene description API, while the low-level rendering API, described in this section, is \textit{egocentric} (i.e. source position and orientation coordinates are defined as "head-relative").

\medskip
\subsubsection{3-band parametric spectral control} \label{sec:globalProps} \hfill 

\begin{figure} [!htbp]
    \centering
    \bigskip
    \medskip
    \includegraphics[width=0.9\columnwidth]{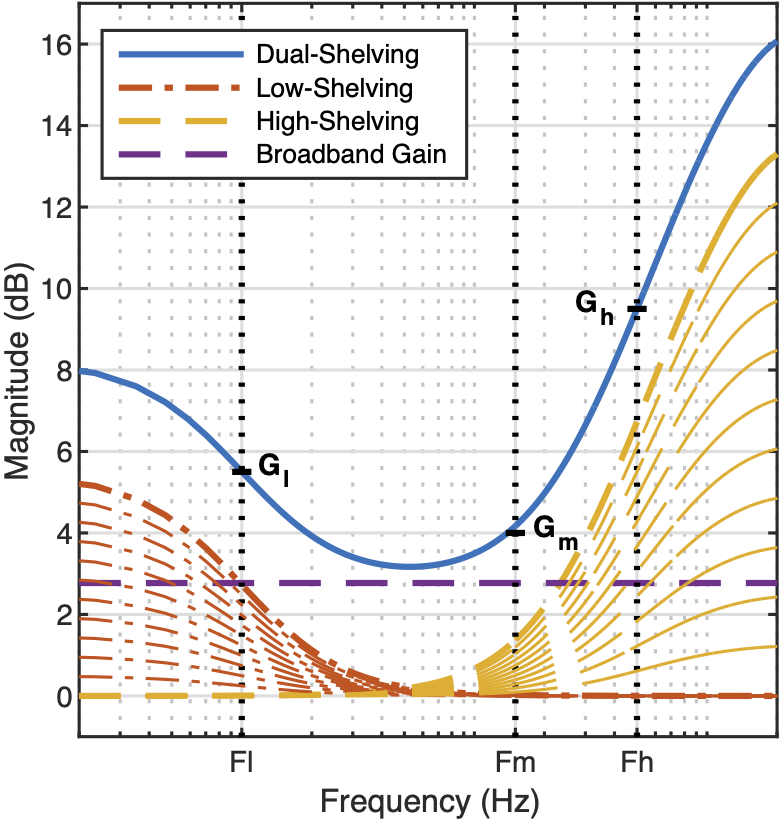}
    \smallskip
    \caption{Parametric magnitude frequency response equalization employing a dual-shelving equalizer with three adjustable control frequencies $(Fl,Fm,Fh)$, realized by cascading two first-order proportional shelving equalizers into a second-order biquad IIR filter module \cite{jot1995spatreference, jot2015proportional, audfray2018dsf}.}
    \label{fig:BLS_EQ}
\end{figure}

Three control frequencies $(Fl,Fm,Fh)$ are specified globally and enforced throughout the audio rendering API to set or calculate frequency-dependent properties and parameters. This approach, inherited from Ircam Spat, I3DL2 and EAX, exploits the proportionality property of the shelving equalizer described in \cite{jot2015proportional} and Fig. \ref{fig:BLS_EQ}, which allows lumping into one second-order IIR filter Section ("biquad") the combination of several cascaded dual-shelving equalizer effects sharing a common set of control frequencies \cite{jot1995spatreference, jot2015proportional, audfray2018dsf}. This enables the computationally efficient implementation of global and per-source gain corrections in three frequency bands throughout the MSA rendering engine.

In MSA, unlike I3DL2 and EAX, acoustic obstruction and occlusion effects are driven by the high-level API (Section \ref{sec:propagation}), whereas the low-level API only exposes rendering filter gain offsets (which can also be used by the programmer for other purposes, including dynamic creative effects) \cite{mlsdk, audfray2018audio}.

\newpage
\subsubsection{Frequency-dependent source directivity} \label{sec:directivitymodel} \hfill 

As in OpenAL EFX, the $Outer\_gain$ property of a source is frequency-dependent so that the direct sound component will be automatically filtered according to the orientation of the source relative to the position of the listener, resulting in a natural sounding dynamic low-pass filtering effect in accordance with listener navigation or when the source points away from the listener.

Additionally, a dual-shelving filter approximating the \textit{diffuse-field transfer function} of the sound source is derived and applied to its reflections and reverberation. As a result, simulating a source that is more directive at higher frequencies, as is typical of natural sound sources, automatically produces a low-pass spectral correction of its reverberation. This is a noticeable timbral characteristic specific to each sound source according to its directivity vs. frequency, predicted by the statistical model developed in \cite{jot1997analysis}. (In Ircam Spat, this natural property of a sound source can be tuned directly by adjusting its '$Omni$' dual-shelving filter parameters \cite{jot1995spatreference, carpentier2015spat}).

\medskip
\subsubsection{Clustered reflections spatialization}
\label{sec:clusterdef}
\hfill 

MSA includes a method for efficient control and rendering of the early reflections, originally proposed in \cite{jot2006binaural}, using a shared \textit{clustered reflections} generator (see Fig. \ref{fig:6dofRenderer} and Sections \ref{sec:reverb_rendering} and \ref{sec:reflections} for more details), which enables controlling  separately for each individual sound source the following properties:
\begin{itemize}
    \item $Reflections\_delay$: relative delay of the first reflection received by the listener for this sound source.
    \item $Reflections\_gain$: the gain of the early reflections cluster for each sound source is adjustable in three frequency bands, as described previously.
    \item $Reflections\_pan$: an azimuth angle, relative to the listener's "look direction", that specifies the average direction of arrival of the early reflections cluster for this source.
    \item $Reflections\_focus$: equal to the magnitude of the combined Gerzon Energy Vector of the reflections \cite{gerzon1992general}, represents the directional focus of the clustered reflections around their average direction of arrival (see Fig. \ref{fig:diffusepanning}) -- with maximum directional concentration for a value of 1.0. For a value of 0.0, the incidence of the reflections is evenly distributed (isotropic) around the listener's head.
\end{itemize}

 By default, the clustered early reflections are spatialized to emanate from a direction sector centered on the direct sound arrival for each source (as illustrated in Fig. \ref{fig:diffusepanning}). As in Ircam Spat, this behavior supports the auditory perception of source localization without requiring geometrical and acoustical environment data \cite{jullien1995, jot1999acm}. Alternatively, in VR or AR applications where a physical environment description is accessible, the clustered reflections properties listed above may be controlled programmatically for each sound source (see Section \ref{sec:propagation}).

\newpage
\subsubsection{Enhanced distance model} \label{sec:distancemodel} \hfill 

\begin{figure} [t] 
    \centering
    \includegraphics[width=0.95\columnwidth]{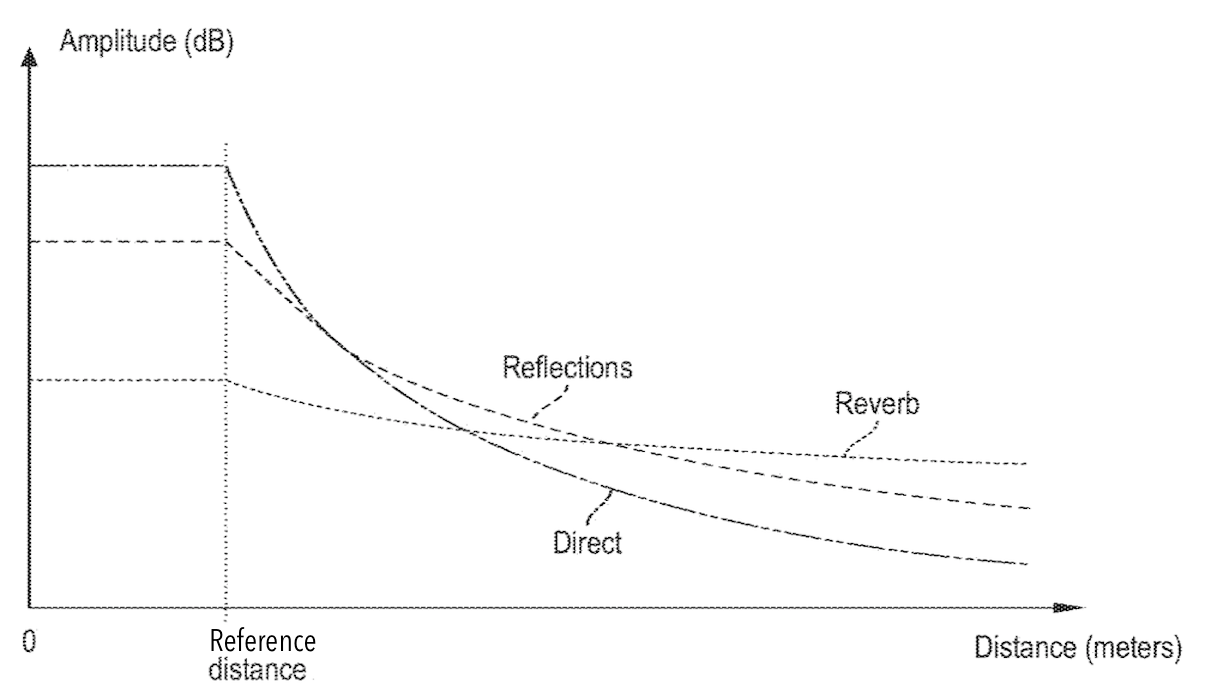}
    \caption{Example of automatic attenuation vs. distance differentiated for the direct sound, the reflections and the reverberation.}
    \label{fig:gainVSdist}
\end{figure}

As illustrated in Fig. \ref{fig:gainVSdist}, the automatic attenuation of sound components vs. distance is extended to manage separately the direct sound, the early reflections and the late reverberation intensities for each sound source. For an unobstructed sound source located beyond the $Reference\_distance$ in the same room as the listener, the following behavior applies by default:
\begin{itemize}
    \item The direct sound component is subject to the conventional attenuation behavior, referred to as 'inverse distance clamped' model in OpenAL \cite{OpenAL, i3dl1}.
    \item The early reflections are subject to the same relative attenuation as the direct sound component. This emulates the perceptual model realized in Ircam Spat, wherein the attenuation applied to the early reflections segment substantially matches the attenuation applied to the direct sound as the source distance varies \cite{jullien1995, jot1995spatreference, jot1999acm}.
    \item The late reverberation is subject to a frequency-dependent attenuation according to the room's reverberation decay time, as explained in Section \ref{sec:reverbfingerprint} below.
\end{itemize}

\medskip
\subsubsection{The reverberation fingerprint of a room}  \label{sec:reverbfingerprint} \hfill 

Following an acoustic pulse excitation in a room, the reflected sound field builds up to a \textit{mixing time} after which the residual acoustic reverberation energy is distributed uniformly across the enclosure \cite{jot1997analysis, polack1993}. A numerical simulation example included in Appendix illustrates this phenomenon. The ensuing reverberation decay can be simulated accurately for VR or AR applications by considering, regardless of source or listener position and orientation \cite{jot2016augmented, jot1997analysis}:
\begin{itemize}
    \item an intrinsic property of the room: its \textit{reverberation fingerprint}, encompassing the reverberation decay time (function of frequency) and the room's cubic volume, through which the source's acoustic power is scattered
    \item an intrinsic property of the sound source: its \textit{diffuse-field transfer function} (see Section \ref{sec:directivitymodel}), which scales the acoustic power radiated into the room, accumulated over all directions of emission.
\end{itemize}

\newpage

\begin{figure} [t] 
    \centering
    \includegraphics[width=0.9\columnwidth]{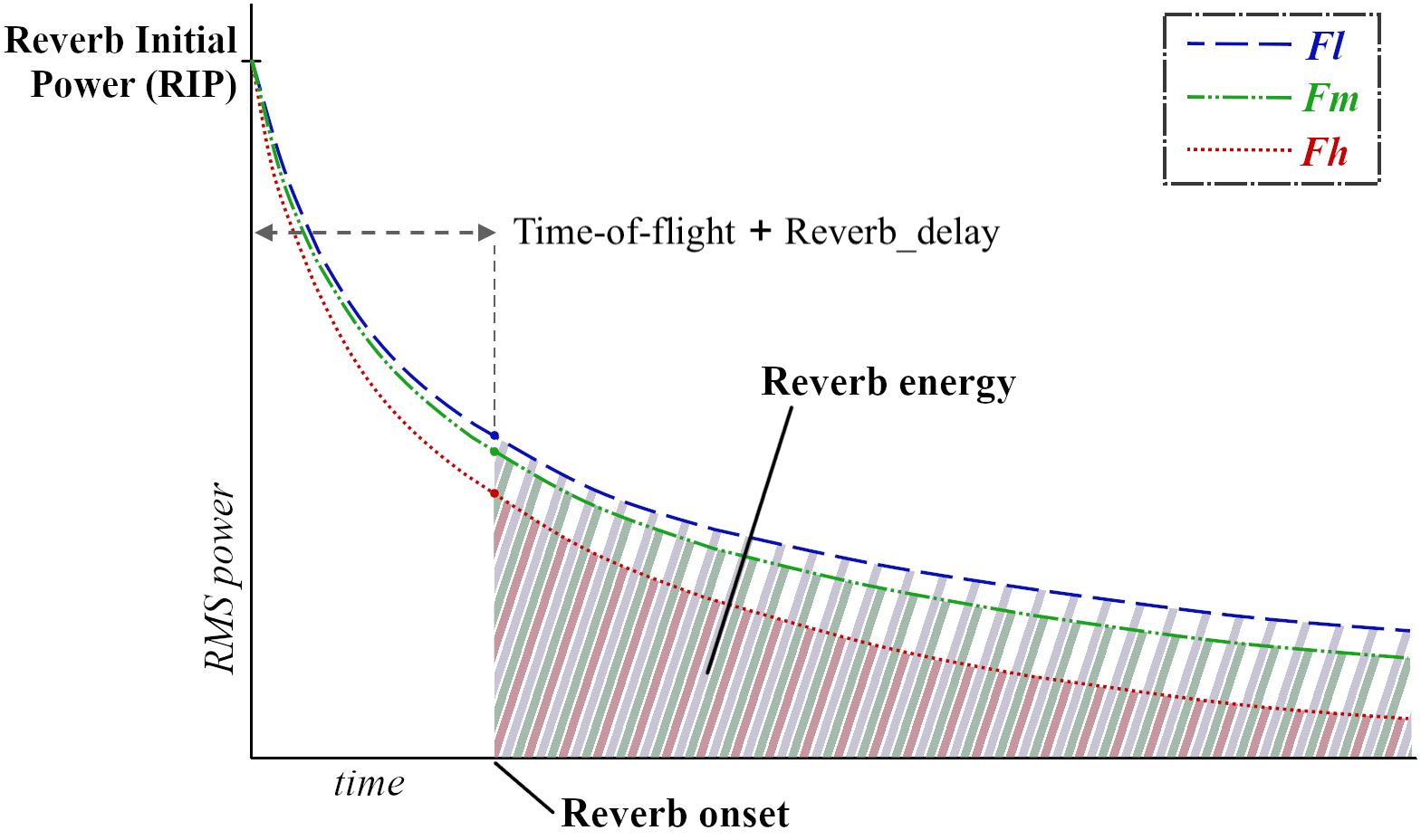}
    \caption{Illustration of energy calculation yielding the relative Reverb attenuation vs. source-listener distance according to the reverberation decay time at three control frequencies $(Fl,Fm,Fh)$.}
    \label{fig:reverbEnergy}
\end{figure}

The distance model described previously in Section \ref{sec:distancemodel} requires the rendering engine to attenuate the reverberation with a per-source frequency-dependent offset that is a function of the source-listener distance and depends on the reverberation decay time, as illustrated in Fig. \ref{fig:reverbEnergy}. This offset accounts for the variation of the remaining integrated energy under the reverberation's power envelope after an onset time equal to the source-to-listener time-of-flight augmented by the $Reverb\_delay$. The $Reverb\_gain$ property value can be thought of as representing the acoustic power amplification by the virtual room's reverberation to a sound generated by the user, such as one's own voice or footsteps \cite{audfray2019reverberation}. 

\subsection{Summary: Audio Scene Rendering for the Metaverse} 

We have presented a generic egocentric rendering API for interactive 6DoF spatial audio and multi-room reverberation, emphasizing auditory plausibility and computational efficiency, compatible with both physical and perceptual scene description models, along the principles proposed previously in \cite{jot-trivi2006}. Control via a perceptual audio spatialization paradigm is illustrated in \cite{trivi-jot2002}. Mapping from a geometrical and physical scene model will be discussed in Section \ref{sec:propagation}.

In particular, we extend the OpenAL EFX feature set to enable per-source spatialization of the early reflections by the \textit{clustered reflections} rendering method previously envisioned in \cite{jot2006binaural} and detailed further in Section \ref{sec:rendering_algorithms}. This enables simulating the perceptually-based distance effect afforded by Ircam Spat's  \textit{Source Presence} parameter \cite{jullien1995, jot1995spatreference, jot1999acm}.

For AR and VR application, the Audio Metaverse is modelled as a collection of rooms each characterized by a "reverberation preset" representative of its response to an omnidirectional sound source located near the receiver. We exploit the notion of \textit{reverberation fingerprint}, which provides a data-efficient characterization of the perceptually relevant characteristics of a room's reverberation that are independent of source or receiver parameters \cite{jot2016augmented}. Virtual sound sources represented by their directivity properties can be seamlessly "dropped" into the environment at rendering time as audio objects assigned arbitrary dynamic position and orientation.

The reverberation fingerprints of the rooms composing a sector of the Metaverse may be inventoried for future retrieval and rendering, and estimated by automated acoustic analysis \cite{murgai2017blind, gamper2018blind}. A future extension is the validation and extension of this model for "non-mixing" or anisotropic reverberation environments such as open or semi-open spaces, coupled rooms, or corridors \cite{alary2021}. Future work also includes incorporating the capability of representing spatially extended sound sources in this rendering API \cite{potard2004, verron2010}.

\bigskip
\section{A Binaural Immersive Rendering Engine} \label{sec:rendering_algorithms}

\begin{figure} [!htbp]
    \centering
    \medskip
    \includegraphics[width=\columnwidth]{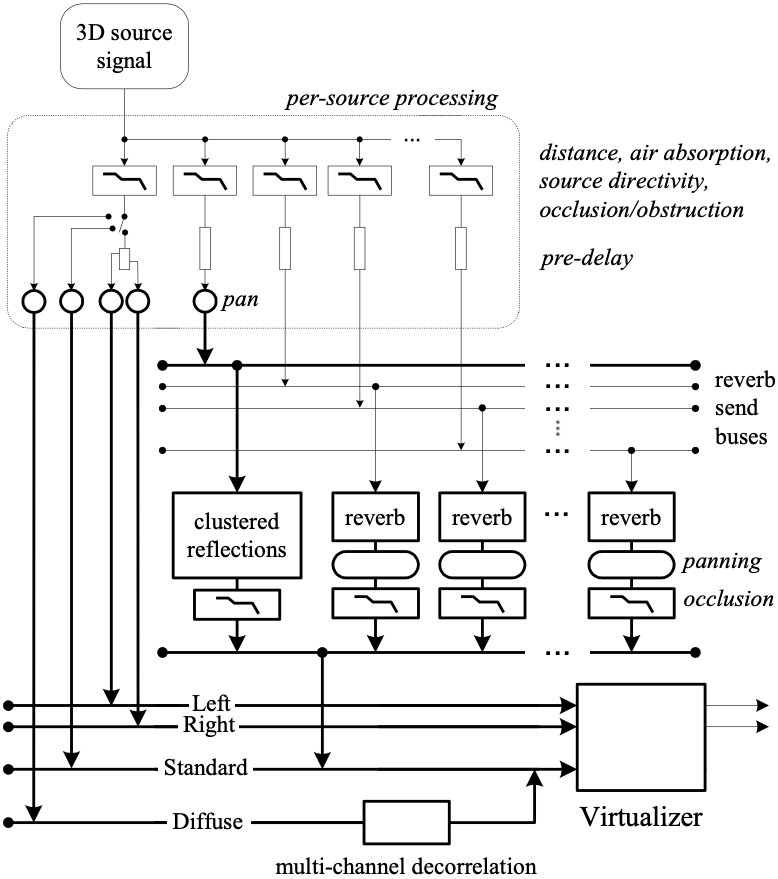}
    \caption{Architecture overview of a 6DoF object-based audio rendering engine including per-source and global parametric spectral correctors \cite{jot2006binaural}. Thicker lines indicate multi-channel audio signal paths or processing functions.}
    \label{fig:6dofRenderer}
\end{figure}

Fig. \ref{fig:6dofRenderer} displays an audio processing architecture that supports the rendering API functions reviewed above. It indicates where dual-shelving equalizers are inserted in order to realize per-source or global spectral corrections (see Section \ref{sec:globalProps}). In this section, we build this audio renderer in successive stages of extension, following the evolution outlined in Table \ref{table:APIcomparison}:
\begin{itemize}
    \item 3D positional direct sound rendering for multiple audio objects, including spatially extended sound sources
    \item spatially diffuse rendering of the late reverberation in the listener's room (Section \ref{sec:reverb_rendering})
    \item addition of the acoustic reverberation emanating from adjacent rooms (Section \ref{sec:multiroom})
    \item computationally efficient rendering of the early reflections for each sound source (Section \ref{sec:reflections}).\\
\end{itemize}

\subsection{Binaural 3D Positional Audio Rendering}

\begin{figure} [!htbp]
    \centering
    \medskip
    \includegraphics[width=0.85\columnwidth]{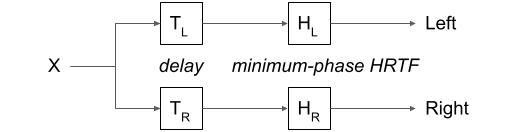}
    \caption{Reference rendering model for one point source.}
    \label{fig:brutForce}
\end{figure}

\begin{figure} [!htbp]
    \centering
    \includegraphics[width=.7\columnwidth]{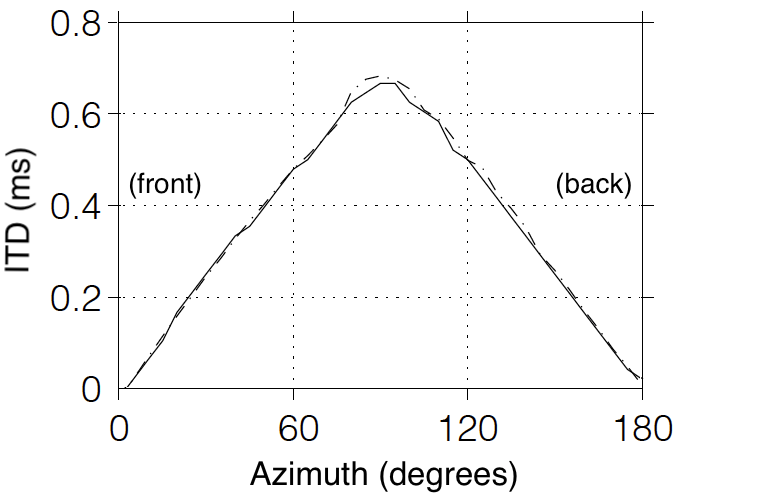}
    \caption{Interaural Time Delay (ITD) derived from the interaural excess-phase difference in HRTF measurements collected on a Head Acoustics HMSII artificial head (after \cite{larcher2001}).}
    \label{fig:itdVSaz}
\end{figure}

In this section, we address the binaural rendering of the direct sound component for a collection of audio objects, critical for faithful reproduction of their individual location, loudness and timbre \cite{blauert1983}. Fig. \ref{fig:brutForce} shows the reference Head-Related Transfer Function (HRTF) filter model for one source at a given location: frequency-independent bilateral delays cascaded with a minimum-phase filter stage \cite{jot1995digital, larcher2001}. Figures \ref{fig:itdVSaz} and \ref{fig:BLS_Virtual_Speakers} display HRTF filter specification examples based on measurements performed on artificial heads. We focus on efficient methods applicable for virtually unlimited object counts, where per-source computation is minimized by (a) panning and mixing in an intermediate multichannel format, then (b) transcoding to binaural output (see e.g. \cite{jot-noh2017}). Fig. \ref{fig:6dofRenderer} distinguishes three multichannel mixing modes:

\medskip
\subsubsection{Standard}
Travis \cite{travis1996} proposed ambisonic to binaural transcoding for interactive audio, including 3DoF compensation of listener head movements by rotation of the B-Format mix. The Virtualizer performs ambisonic decoding to a chosen head-locked virtual loudspeaker layout, as illustrated in Fig. \ref{fig:BLS_Virtual_Speakers}. This approach is extended to high-order ambisonic encoding in \cite{noisternig2003}. Alternatively, a pairwise amplitude panning technique such as VBAP \cite{pulkki1997} may be used, wherein listener head rotations are accounted for at the per-source panning stage. For practical channel counts, these approaches are inaccurate in terms of HRTF reconstruction, which can result in perceived localization errors \cite{jot1999conf, wiggins2017, ben-hur2021}. Frequency-domain parametric processing can mitigate this drawback, at the cost of increasing virtualizer complexity \cite{goodwin2007, breebart2008, berge2010}. 


\medskip
\subsubsection{Left, Right (bilateral)}
\begin{figure} [!htbp]
    \centering
    \medskip
    \includegraphics[width=0.9\columnwidth]{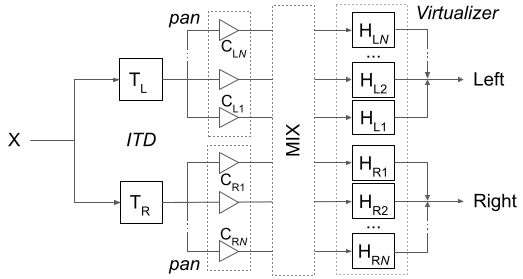}
    \caption{Bilateral multichannel rendering signal flow \cite{jot1999conf, larcher2001}.}
    \label{fig:bilateralMultichan}
\end{figure}

This approach realizes direct ITD synthesis for each individual sound source, as illustrated in Fig. \ref{fig:bilateralMultichan}. It is equivalent to the linear decomposition of the minimum-phase HRTF filter using a basis of spatial panning functions \cite{jot1999conf, larcher2001} -- for instance: first-order Ambisonics \cite{jot1998, jot1999conf}, generalization to spherical harmonics of any order \cite{ben-hur2021}, or bilateral pairwise amplitude panning \cite{jot2006binaural}. These methods can be readily extended to include per-source processing for near-field positional audio rendering \cite{romblom2008nearfield}, and to customize ITD, spatial or spectral functions in order to match an individual listener's HRTF dataset \cite{jot1999conf, larcher2001}.


\medskip
\subsubsection{Diffuse}
This third approach, introduced in \cite{jot2006binaural}, allows rendering spatially extended sound sources (see \cite{potard2004}). Fig. \ref{fig:6dofRenderer} assumes that an identical pairwise amplitude panning method is employed for both the \textit{Standard} mix and the \textit{Diffuse} mix, the latter subjected to a decorrelation processing filter bank prior to virtualization (see e.g. \cite{kendall1995, boueri2004}).

\medskip

\begin{figure} [!htbp]
    \centering
    \includegraphics[width=\columnwidth]{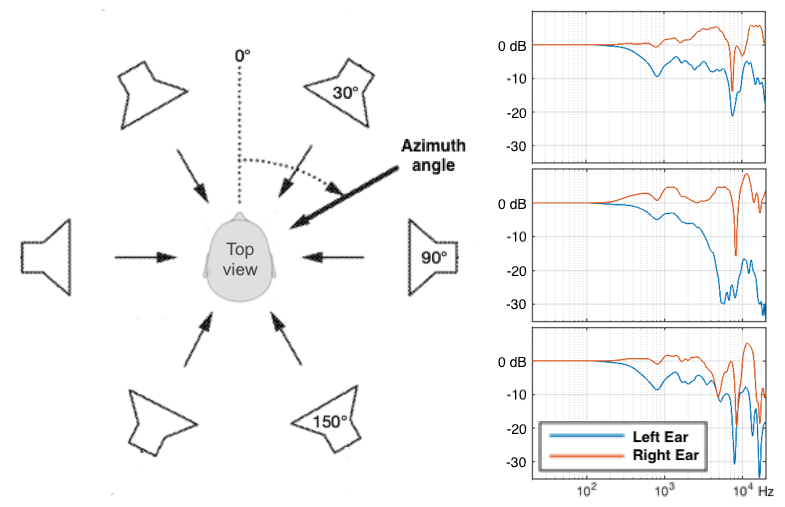}
    \caption{Virtual loudspeakers rendering scheme using 6 horizontal directions, and corresponding HRTF filter magnitude frequency responses for subject 'KU100' in the SADIE II database \cite{sadieII2018} (including corrections for diffuse-field equalization \cite{larcher1998equalization} and low-frequency gain normalization).}
    \label{fig:BLS_Virtual_Speakers}
\end{figure}

\newpage
\subsection{Simulating Natural Acoustic Reverberation}
\label{sec:reverb_rendering}

To incorporate environmental acoustics in the interactive audio rendering engine, EAX and I3DL2 adopted a processing architecture analogous to music workstations, where a reverberator operates as an auxiliary effect that can be shared between multiple sound sources -- except for one important difference: here, the reverberator input filter can be tuned individually for each source according to its diffuse-field transfer function (see Sections \ref{sec:reverbAPI}, \ref{sec:directivitymodel} and Fig. \ref{fig:6dofRenderer}).

\newpage
Reverberation engine designs have been proposed wherein a per-source early reflections generator feeds into a late reverberation processor that can be shared between multiple sound sources located in the same room as the listener \cite{jot1997icmc, desena2015sdn}. Here, as shown in Fig. \ref{fig:6dofRenderer}, the reflections generator is shared too (\textit{clustered reflections} module, Section \ref{sec:reflections}) and the late reverberation is rendered by a \textit{reverb} module which receives a separate mix of the source signals.

\begin{figure} [!htbp]
    \centering
    \includegraphics[width=0.6\columnwidth]{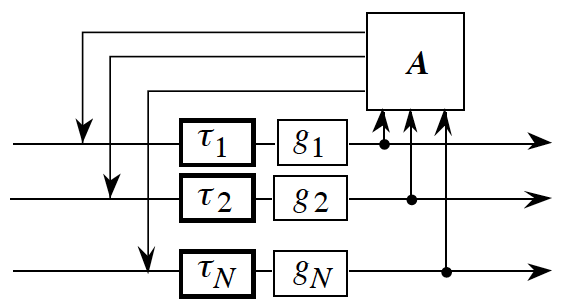}
    \caption{Canonic Feedback Delay Network (FDN) representation. $\{\tau_i,g_i\}$ is a set of absorptive delay units coupled by feedback matrix $A$ (from \cite{jot1997icmc}).}
    \label{fig:fdnReverb}
\end{figure}

    
Parametric reverberation algorithms suitable for this purpose have been extensively studied, including methods for analyzing, simulating or "sculpting" recorded, simulated or calculated reverberation responses \cite{jot1997icmc, gardner2002, valimaki2012fifty, desena2015sdn, carpentier2013, xiang2019, fagerstrom2020vfdn}. Many involve a recirculating delay network as illustrated in Fig. \ref{fig:fdnReverb} -- where, referring to the reverberation API properties introduced in Section \ref{sec:reverbAPI} \cite{jot1997icmc, jot1991digital, gardner2002},
\begin{itemize}
    \item $\{\tau_i\}$ is a set of recirculating delay units, whose summed length may be scaled via the modal $Density$ property
    \item $A$ denotes a matrix of feedback coefficients, whose sparsity may be adjusted via the $Diffusion$ property
    \item each $g_i$ denotes an absorptive filter realizing a frequency-dependent dB attenuation proportional to the length of the delay $\tau_i$ and to the reciprocal of the decay time. 
\end{itemize}

\begin{figure} [!htbp]
    \centering
    \includegraphics[width=0.8\columnwidth]{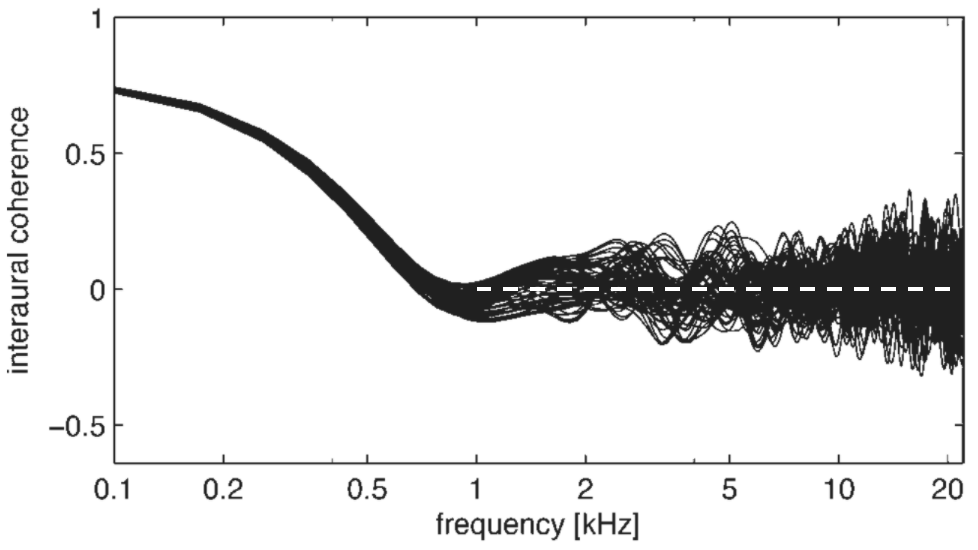}
    \caption{Measured frequency-dependent interaural coherence in diffuse reverberation \cite{menzer2010coherence} and substitution above 1kHz.}
    \label{fig:menzerCoherence}
\end{figure}

These filters may be realized with proportional parametric equalizers as described in \cite{jot2015proportional} and Section \ref{sec:globalProps}, such that the property $Decay\_hf\_ratio$ (resp. $-\_lf\_ratio$) sets the decay time at $Fh$ (resp. $Fl$) relative to the mid-frequency $Decay\_time$. Additionally, cascaded \textit{reverb} and \textit{Virtualizer} processing is normalized to match the $Reverb\_gain$ \cite{audfray2019reverberation} and mimic the low-frequency interaural coherence contour observed in natural diffuse fields (Fig. \ref{fig:menzerCoherence}) \cite{jot1995digital, menzer2010coherence, xiang2019}.

\bigskip

\newpage
\subsection{Multi-Room Reverberation Rendering} \label{sec:multiroom}

As illustrated in Figures \ref{fig:eax4signal} and \ref{fig:6dofRenderer}, several reverberators can be employed to simulate neighboring rooms in addition to an occupied room.  Rendering multiple rooms requires both additional logic governing the amount of signal sent from audio sources to different reverberation processing units as well as a more complicated mechanism for handling multiple reverberation units’ output.

Any source-specific qualities desired in the reverberation output must be imparted when signal is sent from the source to a reverberation processing unit. Existing API solutions employ a reverb send module that sets gain, equalization, and delay for signal sent from a source to a specific reverberation unit \cite{OpenALEFX, eax4programmer, jot2006binaural}. A significant runtime complexity of a multi-room system stems from determining desired values for these source sends, which may simply be based on source position relative to the listener and room geometry but could also account for source orientation and radiation pattern, particularly relevant if a source is positioned near a boundary between rooms.  Considerations such as these, that have been left to the caller by historical APIs, are discussed in Section \ref{sec:propagation}.

\begin{figure} [!htbp]
    \centering
    \bigskip
    \includegraphics[width=\columnwidth]{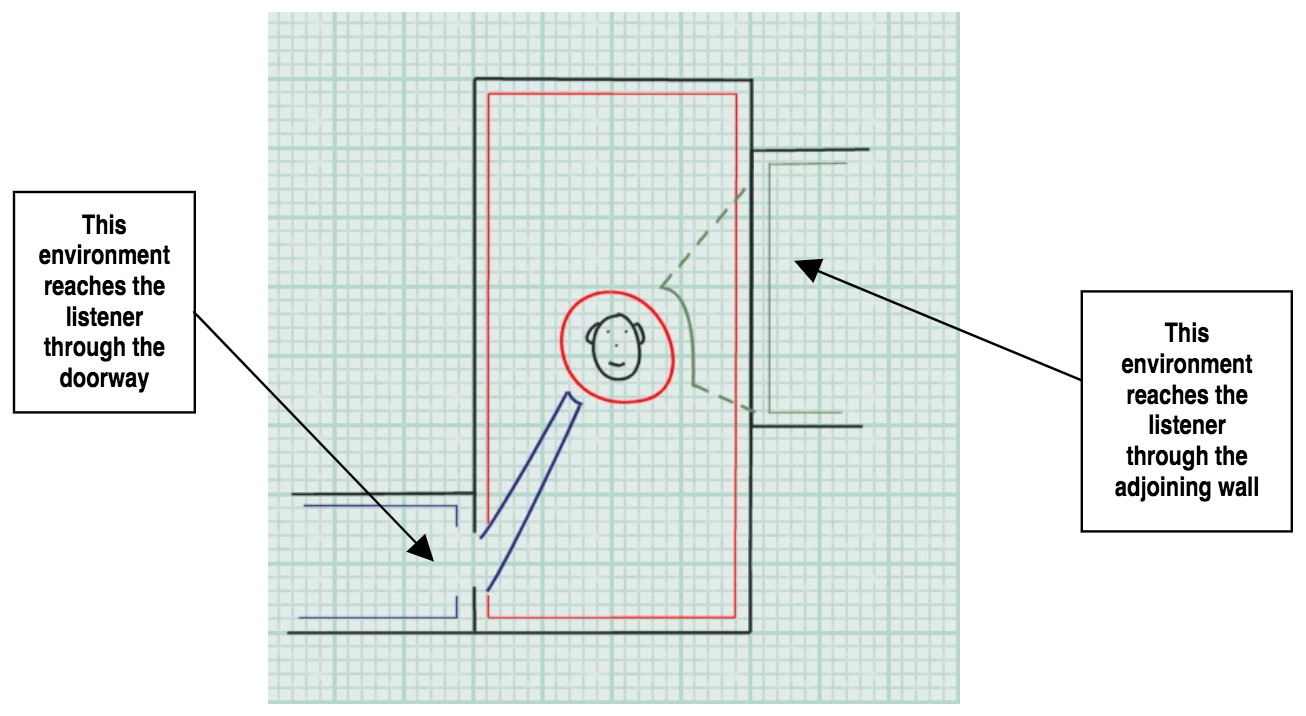}
    \caption{Environmental panning scene example (from \cite{eax4introduction}).}
    \label{fig:efx_extension_env_panning}
\end{figure}

In the case of a multi-room system that renders adjacent rooms, some method of spatially panning reverberation output becomes necessary, as illustrated in Fig. \ref{fig:efx_extension_env_panning}.  As proposed in \cite{jot2006binaural} and shown in Fig. \ref{fig:6dofRenderer}, spatial reverberation panning may be achieved by routing reverberator output through a virtualization algorithm, requiring a number of sufficiently decorrelated reverberation output channels determined by the desired spatial resolution for reverberation output panning.  Alternatively, a dedicated bus whose channels all undergo decorrelation processing prior to virtualization, labeled `Diffuse' in \cite{jot2006binaural} and Fig. \ref{fig:6dofRenderer}, can be employed.  In any case, the spatial resolution of reverb panning cannot exceed that provided by the number and placement of virtual loudspeakers used to render spatial content.

Some representation of a desired spatial panning result is required by the API, ideally one that minimizes its data footprint without compromising signal processing capability; as such it need not specify data beyond the spatial audio rendering system’s ability to translate into perceivable difference. 

\newpage
The EAX Reverb effect allows the setting of direction and dispersion (a measure of how wide or narrow the diffuse sound should be) to pan early reflections and late reverberation (see Fig. \ref{fig:diffusepanning}) \cite{OpenALEFX}.  These two properties together describe a Gerzon vector as discussed in \cite{jot2006binaural}, and on which the MSA source $Reflections\_pan$ and $Reflections\_focus$ properties, used to pan clustered reflections for a source, are similarly based (as discussed in Section \ref{sec:clusterdef}).

The desired spatial resolution for reverberation panning must also be considered by this panning representation, whether by one straightforward such as the one above or by one more sophisticated. Spherical harmonics are a potential candidate for higher resolutions and more complex patterns, with the benefits that they are virtual speaker agnostic, can vary in order, and can render at lower orders at runtime with minimal audio disruption if resources are constrained.  Direct representation of virtual speaker gains is another option, which while dependent on prior knowledge of the virtual speaker configuration in use could leverage standardized positions for virtual speakers, or subsets of these.  

Non-occupied rooms also require an additional equalization stage to apply attenuation as a result of distance and the manner of each room's connection to the occupied room.  For example, one may wish to render bleed of a neighboring room through a wall (i.e. occluded reverberation) as in Fig. \ref{fig:efx_extension_env_panning}.  In \cite{OpenALEFX}, an equalization filter can be applied to reverberation output to simulate these effects; however, the spectral content does not vary across the spatial field.  

One final consideration of a multi-room system is the issue of acoustic coupling between physical rooms the system seeks to model.  Output signal from one or more reverberation units may be cross-fed to the input of one or more other reverberation units to simulate this effect, with the level and equalization of the cross-fed signal for each coupling set via API based on how the physical rooms are connected.

It has historically fallen on the caller of low-level rendering APIs (such as EAX or OpenAL) to determine the proper runtime values for reverb send and reverb output parameters in order to realize an intended multi-room environment, whether bespoke, observed, or some combination of the two. General considerations in this respect will be discussed in Section \ref{sec:propagation}.

\subsection{Spatialization of Early Reflections} \label{sec:reflections}

In order to realize differentiated early reflections control for each sound source (see Section \ref{sec:clusterdef}), a computationally efficient approach is offered by the \textit{clustered reflections} rendering method proposed in \cite{jot2006binaural}. As shown in Fig. \ref{fig:6dofRenderer}, source signals are panned and mixed into a multichannel reflections send bus which feeds the aggregated contribution of all sound sources into a shared \textit{clustered reflections} processing module.

\begin{figure} [t]
    \centering
    \includegraphics[width=0.65\columnwidth]{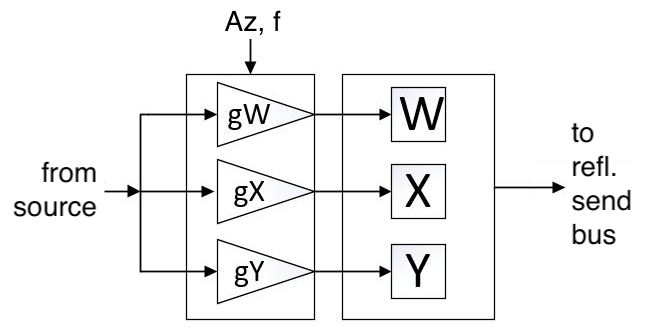}
    \caption{Per-source reflections pan encoder using a 3-channel horizontal Ambisonic signal format.}
    \label{fig:BLS_Reflections_Encoder}
\end{figure}

In this section, we describe a realization in which this send bus employs a 3-channel first-order ambisonic format $(W,X,Y)$. As illustrated in Fig. \ref{fig:BLS_Reflections_Encoder}, encoding is performed with the 2-dimensional ambisonic panning equations \cite{zotter2019}:
\begin{equation}
    g_W = \sqrt{1-f^2},\;\; g_X = f\, cos{Az},\;\; g_Y = f\, sin{Az},
\end{equation}
\noindent where $(g_W,g_X,g_Y)$ is the triplet of panning gains, while $Az$ and $f$ are respectively the per-source $Reflections\_pan$ and $Reflections\_focus$ properties defined in Section \ref{sec:clusterdef}.
 
\begin{figure} [b]
    \medskip
    \centering
    \includegraphics[width=\columnwidth]{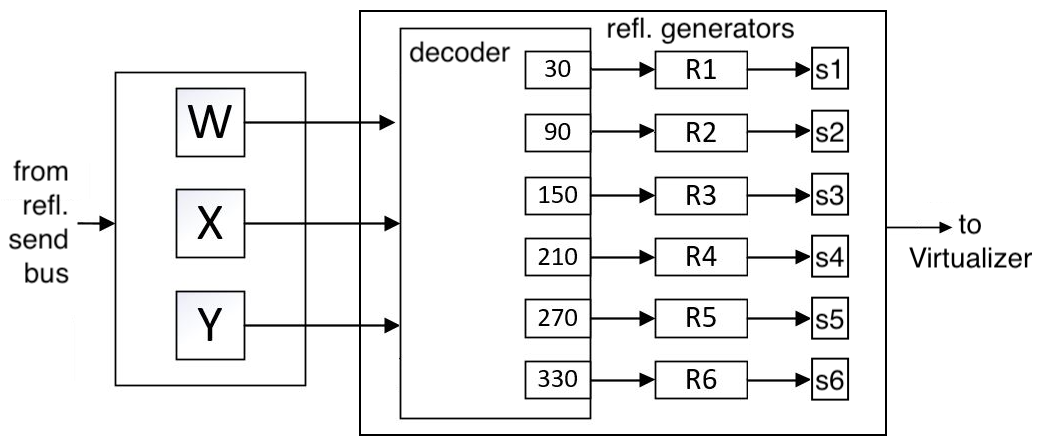}
    \caption{Clustered reflections processing module.}
    \label{fig:BLS_Decoder_Generator}
    \medskip
\end{figure}

As shown in Fig. \ref{fig:BLS_Decoder_Generator}, the reflections send bus signal is decoded into 6 head-locked virtual locations spaced equally around the listener in the horizontal plane (Fig. \ref{fig:BLS_Virtual_Speakers}), by use of the \textit{"in-phase"} ambisonic decoding equations \cite{jot1999conf}:
\begin{equation}
    r_i = \frac{2}{9}\, (W + X\, cos{Az_i} + Y\, sin{Az_i}),
\end{equation}
\noindent for $i$ = 1..6, where $r_i$ denotes a decoded output channel signal and $Az_i$ denotes the corresponding virtual loudspeaker azimuth angle in the layout of Fig. \ref{fig:BLS_Virtual_Speakers}. Each decoded signal is fed into an individual reflections generator realized by a nested all-pass filter \cite{gardner2002} to produce a 6-channel reflections output signal that is forwarded via the Standard multichannel mix bus to the global Virtualizer (Fig. \ref{fig:6dofRenderer}).



\bigskip
\section{Considering Acoustic Propagation} \label{sec:propagation}

When considering support of advanced propagation-related features, one finds existing rendering engines such as MSA and its predecessor OpenAL EFX provide the means to realize an intended perceptual result, such as specifying equalization to be applied to a spatial source to simulate an occlusion effect.  However, the conditions under which the effect should be applied, and the equalization itself, are left to the programmer (see e.g. \cite{OpenALEFX, eax4programmer, eax4introduction}). These APIs prioritize universality, leaving choices regarding how best to utilize the signal processing components exposed by the rendering API to users building a wide range of applications, from the music and media uses discussed in Section \ref{sec:media} to the immersive 6DoF VR or AR experiences encountered in the Metaverse.

\newpage

Complete spatial audio solutions supporting propagation features, including commercially available products such as Audiokinetic's Wwise \cite{wwiseWeb} or Apple's PHASE \cite{applePHASE}, are also faced with determining the desired momentary perceptual result based on position and orientation of a user or avatar within a dynamic acoustic environment.   While reasonable for a lightweight interface prioritizing portability to pass on these determinations to the developer, the emergence and success of more dedicated and feature-rich solutions indicates that with increased familiarity, improved computational capabilities, and larger yet more focused use cases (e.g. the Metaverse), support of higher-level propagation features has become more desired and expected.  This expanding prevalence prompts discussion of terms and principles, such as material properties relevant to different propagation features, for any solution seeking to support these features while maintaining universality.  

Ultimately, we envision a similarly-portable propagation-focused software layer on top of the existing rendering interface described in Section \ref{sec:rendering_api}, encapsulating the related computation needed to drive the perceptual properties of the lower-level interface.  Algorithmic design decisions involving this propagation layer may span a wide range and have implications on how the perceptual layer beneath is employed, adding to the challenge of a universal solution.  A primary role of such a propagation layer, as it seeks to simulate propagation in an acoustic environment, is the management of the acoustic environment itself. The data required to represent the acoustic environment is to some extent determined by algorithmic choices, but vocabulary and representation of perceptually measured properties of individual acoustic objects or elements in the acoustic scene as they relate to propagation features can be more universally considered.  

\subsection{Obstruction}

\begin{figure} [!htbp]
    \centering
    \includegraphics[width=0.75\columnwidth]{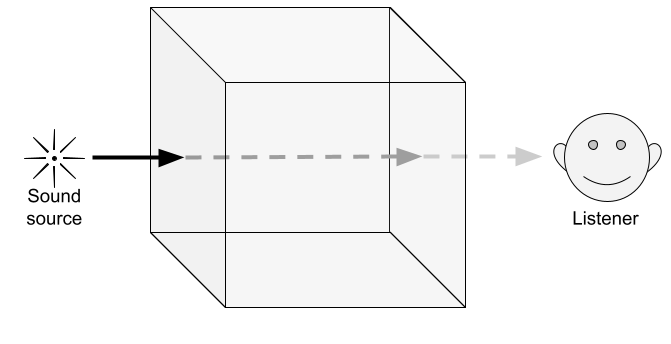}
    \caption{Obstruction - Acoustic object property: $Transmission$.}
    \label{fig:obstruction}
\end{figure}

In considering obstruction of the direct path (Fig. \ref{fig:obstruction}), the acoustic transmission loss through an obstructing object is determined by variances in the acoustic impedance along the sound path as it travels through the object. 
It is convenient to lump the perceptual result of this multi-stage effect (with the assumption the surrounding medium is air near room temperature and atmospheric pressure) into a single set of equalization values that can be stored with the object's acoustic data as a property named $Transmission$. This transmission equalization can then be simply applied to sound paths traveling through the object. The number of equalization bands specified in $Transmission$ should correspond with that used universally throughout the API; in the case of MSA the equalization values correspond to the three-band paradigm discussed in Section \ref{sec:globalProps}. 

\subsection{Reflections}

\begin{figure} [b] 
    \centering
    \includegraphics[width=0.75\columnwidth]{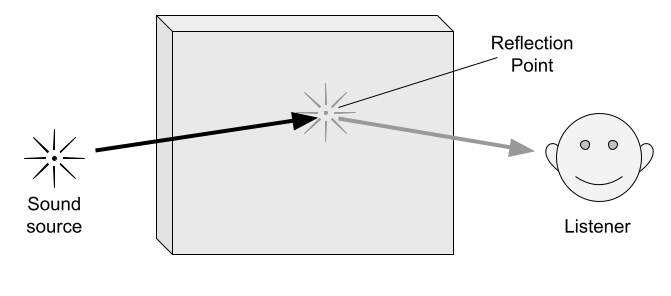}
    \caption{Reflection - Acoustic object property: $Reflectivity$.}
    \label{fig:reflection}
\end{figure}

Relating to objects’ effects on reflections (Fig. \ref{fig:reflection}), one considers the ‘acoustic transmittance’ of a reflecting surface in a similar manner as transmission, with a reflective impedance whose resulting effect we save with objects’ acoustic data under $Reflectivity$.

Put more simply, $Reflectivity$ represents the gain and equalization resulting from a single reflection (or 'bounce') off the object, simplifying computation of higher order reflections’ compounded effects.  Specular and diffuse scattering due to reflection are not separately considered at the object level, although they could be affected by other acoustic object properties, such as geometry or a roughness trait.  $Reflectivity$ may be applied at an equalization stage on either an additional source, modelling a discrete specular reflection of an original source, or it may be applied on a clustered reflections send or output, modelling grouped reflections and having a diffuse effect imparted by the nested all-pass reflections cluster algorithm (see Section \ref{sec:reflections}).  In this latter case, the $Reflectivity$ of multiple nearby objects may be considered in setting a clustered reflections output equalization stage.  It may be noted that transmission, transmittance, and absorption must sum to one, at least for physics-compliant objects.  Because absorption is not quantified, violation of this law can only be identified if $Transmission$ and $Reflectivity$ values sum in excess of one, if such conservation is a priority.  As with other such intention-dependent considerations, enforcement of physically realistic values is left to the caller of the API.

\subsection{Diffraction}

\begin{figure} [t] 
    \centering
    \includegraphics[width=0.75\columnwidth]{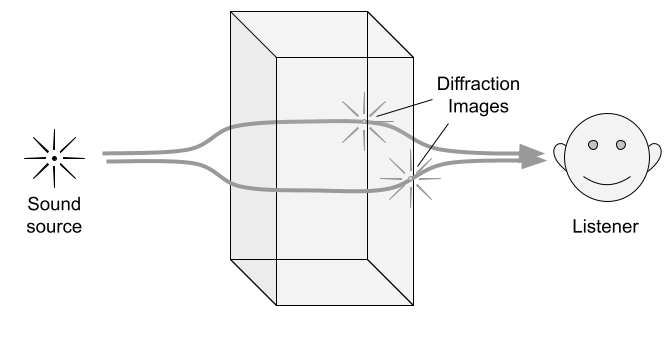}
    \caption{Diffraction - Acoustic object property: Geometry.}
    \label{fig:diffraction}
\end{figure}

Discussion of an acoustic object’s physical properties in the context of diffraction becomes less clear (Fig. \ref{fig:diffraction}).  Acoustic transmittance may play a marginal role, but more important are aspects of geometry, including object shape, size, and orientation. Discretely computed diffraction images, e.g. placed based on shortest-path search around the object at runtime (as in Fig. \ref{fig:diffraction}), may wish to consider the object’s $Reflectivity$ in addition to the original source’s distance attenuation properties when adjusting the image gain and equalization.

Diffraction models that depend on offline computation such as \cite{rungta2018diffraction} may be reliant on whether such computation considers material properties of the acoustic object for variation of these properties to have an audible effect, unless a runtime variable is expressly provided for this purpose. Machine learning based approaches have also been considered that address multiple propagation effects simultaneously \cite{tang2021learning}.

\subsection{Computational Efficiency and Hybrid Architecture}

Any performance-conscious propagation layer, or audio system in general, should have as few variables to set or have to compute at runtime as necessary.  Where possible, psychoacoustic as well as signal processing principles can be exploited to identify and optionally eliminate processing or variables that result in little to no difference in the user experience, guiding design decisions surrounding sufficient yet succinct runtime data representation that weigh fidelity and realism against performance and computational cost.  Existing portable or modular solutions generally rely on perceptual audio / acoustics / psychoacoustics concepts to guide the form of runtime data, both for conciseness and universality, while top-to-bottom solutions can select the form of runtime data most efficiently consumed at runtime within their system.  

Another technique for diminishing runtime demands is to remove variables and computation from runtime operations entirely, and replace them with a compressed representation or other more easily consumed result of some offline computation, whether performed as a dedicated prior step, or as part of a compile or build process, or in a concurrent side process, perhaps on the edge or cloud.  For example, Rungta et al. propose pre-computing a diffraction kernel for individual acoustic objects that considers object properties including geometry \cite{rungta2018diffraction}. This kernel is then used at runtime to compute the momentary transfer function to apply to the direct path based on input and output incident angles. The kernel could be replaced by a lookup table containing values for perceptual properties such as binaural delay, pan direction, and equalization to be set in a lower level rendering engine such as the solution described in Sections \ref{sec:rendering_api} and \ref{sec:rendering_algorithms}.  

Larger scale examples of this strategy that target room or environment simulation also exist.  Microsoft's Project Acoustics simulates room impulse responses using computationally intensive acoustic simulation run on Microsoft's cloud services platform, but from that analysis derives a small number of parameters for runtime use \cite{raghuvanshi2018, chemistruck2020}.  Magic Leap's Soundfield Audio solution (MSA) analyzes captured audio signals in a separate software component to compute reverberation properties, creating an efficient data representation of the acoustic environment (using the reverberation fingerprint discussed in Section \ref{sec:reverbfingerprint}) which may be derived from the analysis of in-situ microphone-captured audio signals \cite{murgai2017blind, gamper2018blind}.








\bigskip
\section{Conclusion}

In combination with emerging wearable augmented reality hardware and with the growing cloud-based ecosystem that underpins the Metaverse, binaural immersive audio rendering technology is an essential enabler of future co-presence, remote collaboration and virtual entertainment experiences. We have presented an efficient object-based 6-degree-of-freedom spatial audio rendering solution for physically-based environment models and perceptually-based musical soundscapes.

In order to support navigable Metaverse experiences where sound source positions and acoustic environment parameters may be served at runtime from several independent applications, the proposed solution facilitates decoupling and deferring to rendering time the specification of the positional coordinates of sound sources and listeners, and of the geometric and acoustic properties of rooms or nearby obstacles and reflectors, real or virtual, that compose the acoustic environment. The design of the renderer and of its programming interface prioritize plausibility (suspension of disbelief) with minimal computational footprint and application complexity. They expose, in particular, these novel capabilities:
\begin{itemize}
    \item The characterization of rooms by their \textit{reverberation fingerprint} enables the faithful matching of acoustic reverberation decays, based on a compact data representation of virtual or real environments.
    \item An efficient method for per-object control and rendering of \textit{clustered reflections}, facilitating perceptually-based distance simulation for each sound source, and avoiding the burden of reproducing early reflections individually in physically-based models.
\end{itemize}

The proposed solution builds upon well established and extensively practiced interactive audio standards, and is implemented as a core software component in Magic Leap's augmented reality operating system.



\bigskip
\section{Acknowledgments}

The authors wish to acknowledge the contribution of Sam Dicker to the development of the rendering engine presented in this paper, as its principal software architect from its beginnings at Creative Labs to its recent maturation at Magic Leap. Jean-Marc would like to acknowledge collaborators in previous work referenced in this paper: Jean-Michel Trivi, Daniel Peacock, Pete Harrison and Garin Hiebert for the development and deployment of the EAX and OpenAL APIs; Olivier Warusfel, Véronique Larcher and Martin Walsh for acoustic and rendering research; Antoine Chaigne for initiating and supporting the investigation as doctoral supervisor.

\newpage

\bibliographystyle{IEEEtran}
\bibliography{main}

\newpage

\appendix[Numerical Simulation of Room Reverberation] \label{sec:RoomSim}

As stated in Section \ref{sec:reverbfingerprint}, the statistical reverberation model developed in \cite{jot1997analysis} leads us to propose the \textit{reverberation fingerprint} as an intrinsic property of a room \cite{jot2016augmented}. Specifically, this model posits that the time-frequency power envelope of the reverberation decay in a (mixing) room is a distinctive characteristic of the room itself (see \cite{jot1997analysis}, Section 4).

By way of practical illustration, the present Appendix provides an example showing that the reverberation decay envelope at a given frequency is independent of source or receiver position. This observation supports the Reverb energy-vs-distance calculation described in Section \ref{sec:reverbfingerprint} and Fig. \ref{fig:reverbEnergy}. We use a commercially available Finite Element Modeling (FEM) tool \cite{comsolweb}. For this illustration, we set up a simple 2D problem with the following conditions:
\begin{itemize}
    \item Omnidirectional point source, Gaussian pulse excitation
    \item Room model dimensions: approx. 5m x 8m.
    \item Broadband wall impedance (identical on all walls)
    \item Time-dependent solver using 0.1 ms steps
    \item Maximum mesh element size: 70cm (making the simulation valid up to about 1kHz).
\end{itemize}

\begin{figure*} [!htbp]
    \centering
    \begin{subfigure}{\myfigwidth}
        \centering
        \includegraphics[width=1.03\textwidth]{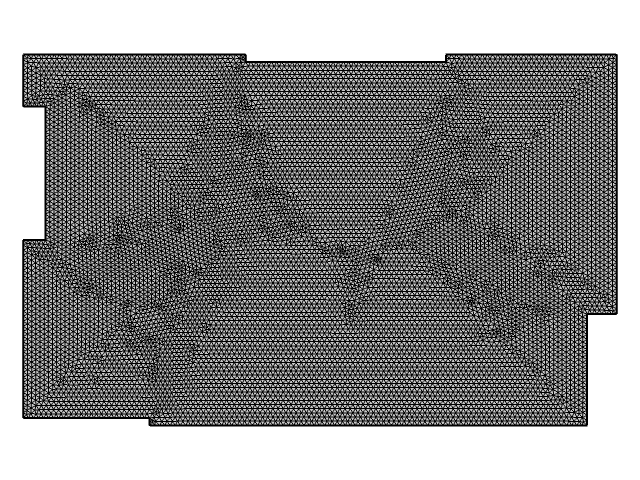}
        \caption{Mesh for FEM}
        \label{fig:roomMesh}
    \end{subfigure}
    \hfill
    \begin{subfigure}{\myfigwidth}
        \centering
        \includegraphics[width=\textwidth]{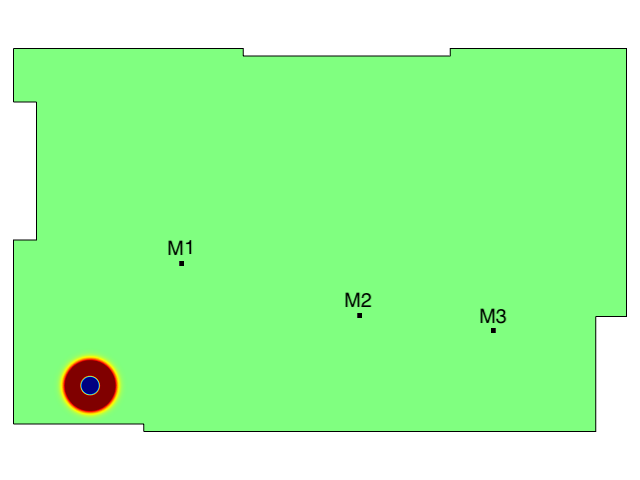}
        \caption{1.5 ms}
    \end{subfigure}
    \hfill
    \begin{subfigure}{\myfigwidth}
        \includegraphics[width=\textwidth]{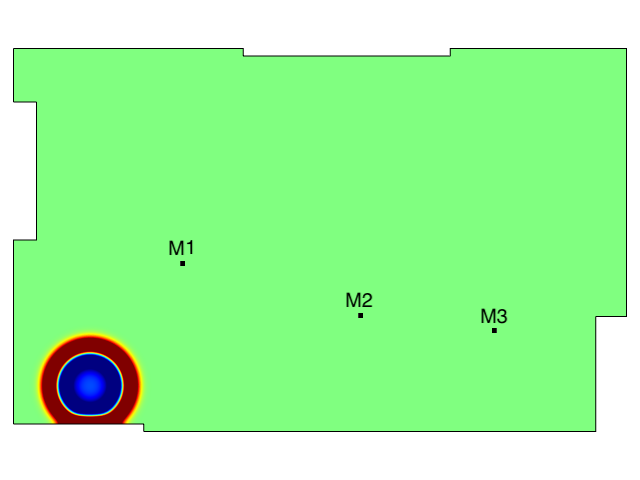}
        \caption{2.4 ms}
    \end{subfigure}
    \hfill
    \begin{subfigure}{\myfigwidth}
        \includegraphics[width=\textwidth]{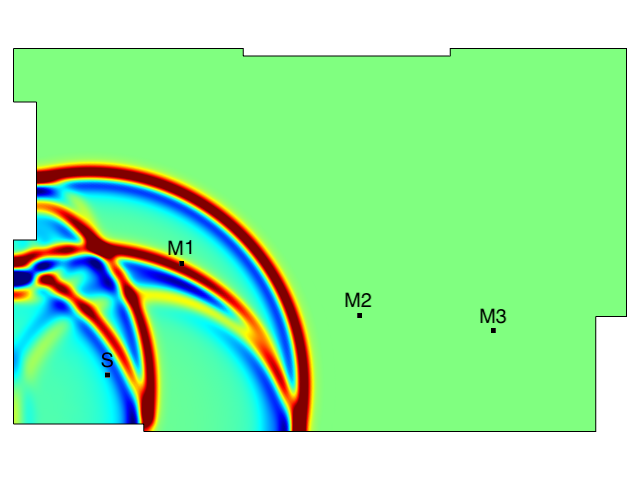}
        \caption{9 ms}
    \end{subfigure}
    \hspace*{\fill}
    \begin{subfigure}{\myfigwidth}
        \includegraphics[width=\textwidth]{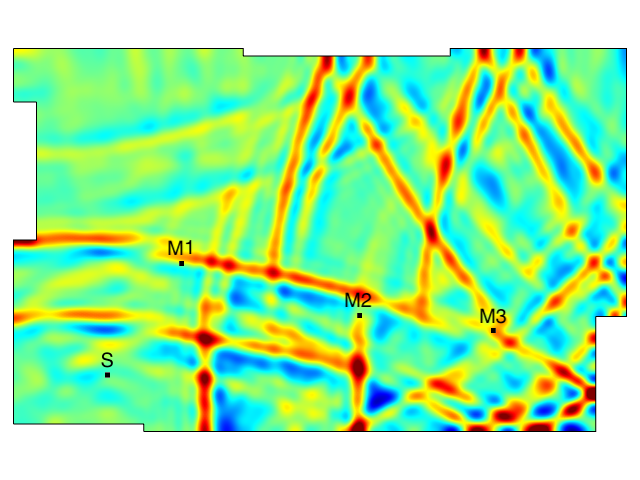}
        \caption{35 ms}
    \end{subfigure}
    \hfill
    \begin{subfigure}{\myfigwidth}
        \includegraphics[width=\textwidth]{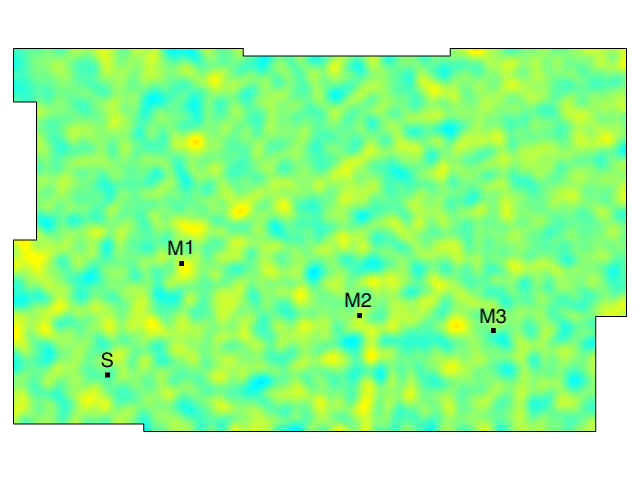}
        \caption{150 ms}
        \label{fig:diffusePressureMap}
    \end{subfigure}
    \hspace*{\fill}
    \caption{Room mesh and calculated pressure maps at different times after emission of an impulse from source position S.}
    \label{fig:pressureMap}
\end{figure*}

Fig. \ref{fig:pressureMap} shows the room model geometry and the locations of a sound source and of three virtual omnidirectional microphones to record room impulse responses at different locations in the simulated space. It also shows the mesh used for the FEM simulation and the evolution over time of the calculated pressure map. It is noticeable that after enough time has passed (the \textit{mixing time} defined in \cite{polack1993}), the reverberant sound field power is uniformly distributed across the room. At 150 ms (Fig. \ref{fig:diffusePressureMap}), it is not possible to guess, by inspection of the pressure field, where the sound source was located.

\begin{figure*} [!htbp]
    \centering
    \hspace*{\fill}
    \begin{subfigure}{\myfigwidth}
        \centering
        \includegraphics[width=\textwidth]{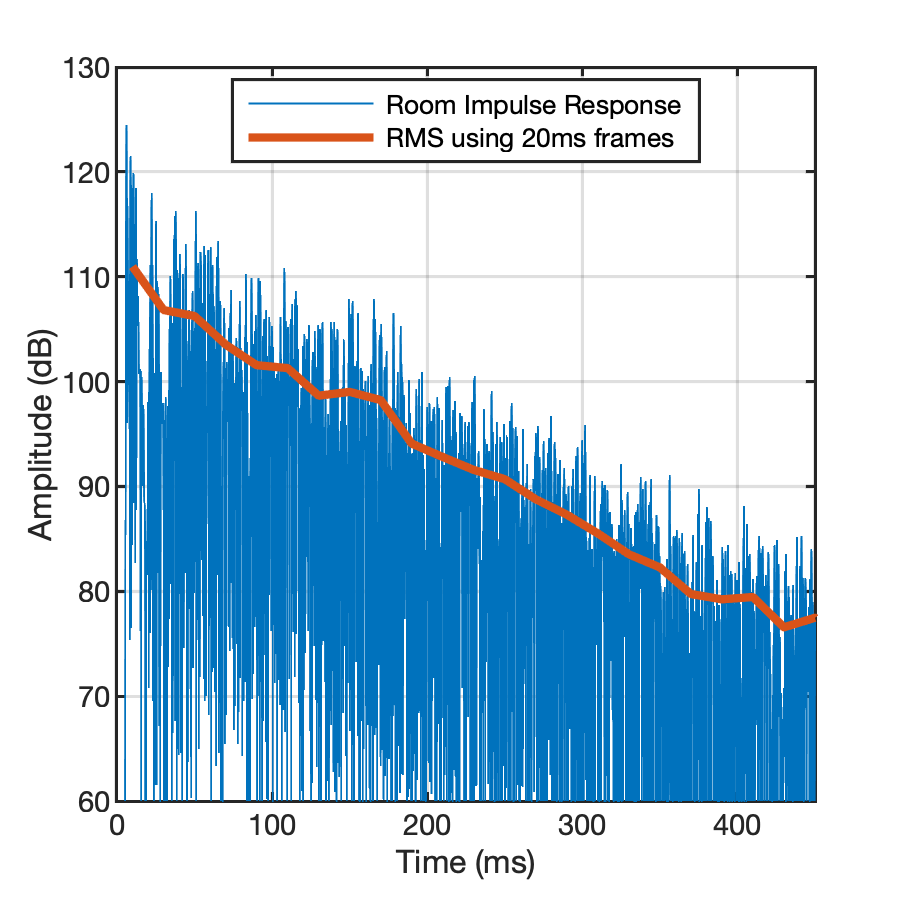}
        \caption{Mic 1} 
    \end{subfigure}
    \hfill
    \begin{subfigure}{\myfigwidth}
        \centering
        \includegraphics[width=\textwidth]{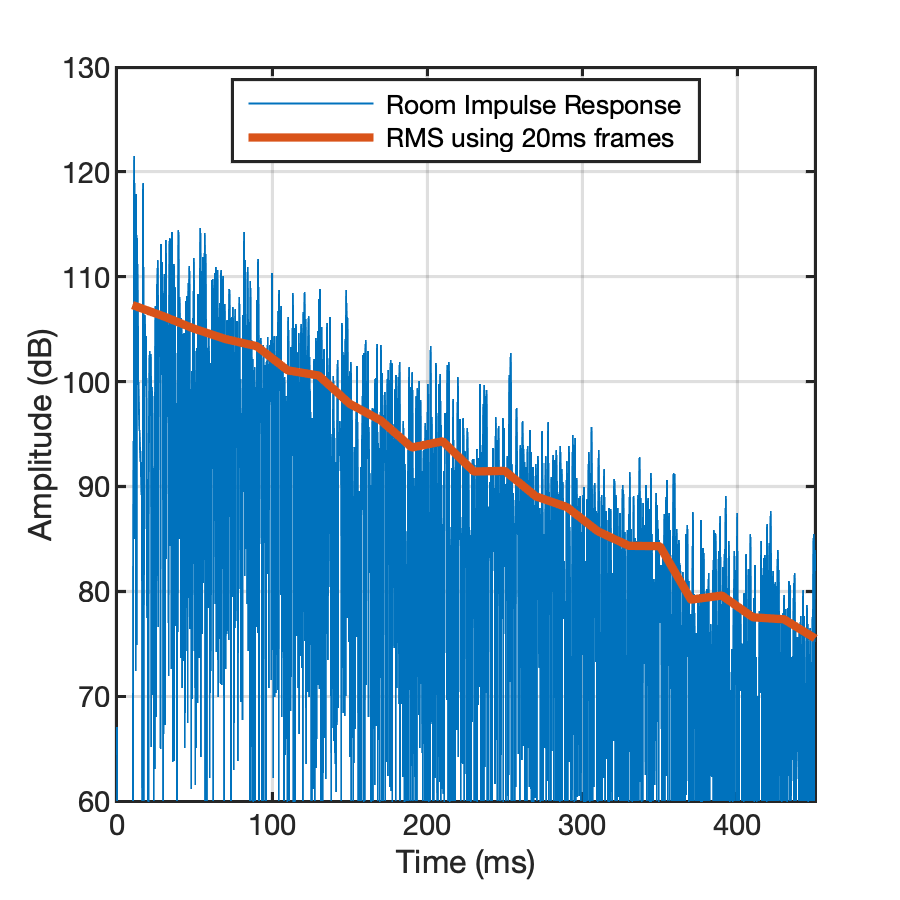}
        \caption{Mic 2} 
    \end{subfigure}
    \hfill
    \begin{subfigure}{\myfigwidth}
        \centering
        \includegraphics[width=\textwidth]{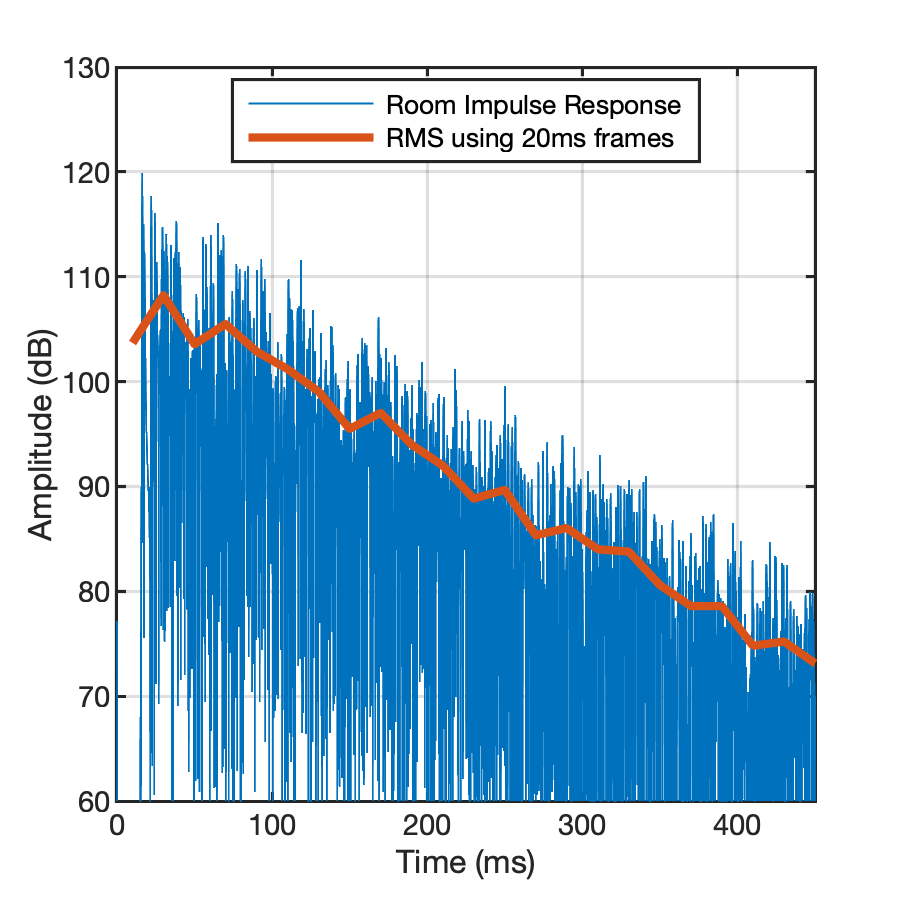}
        \caption{Mic 3} 
    \end{subfigure}
    \hspace*{\fill}
    \caption{Calculated impulse response echograms at microphone positions M1, M2, and M3 shown on Fig. \ref{fig:pressureMap}.}
    \label{fig:roomIRs}
\end{figure*}

\begin{figure*} [b] 
    \centering
    \bigskip
    \includegraphics[width=0.62\columnwidth]{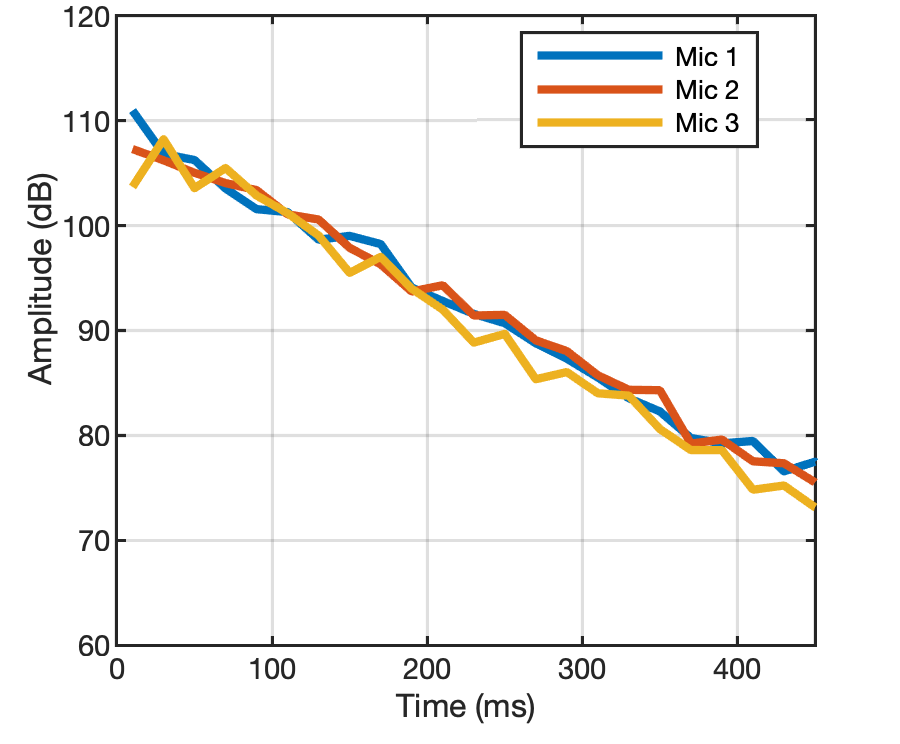}
    \caption{Comparing RMS energy decay at the same three microphone positions as in Figures \ref{fig:pressureMap} and \ref{fig:roomIRs}.}
    \label{fig:roomIRsRMS}
\end{figure*}

Fig. \ref{fig:roomIRs} displays the amplitude of the simulated impulse responses captured by each of the three virtual microphones, as well as their RMS envelopes (also overlaid in Fig. \ref{fig:roomIRsRMS} to facilitate comparison). It is visible that the initial acoustic response is substantially dependent on position, in terms of both direct sound arrival time and early reflection pattern. However, past this initial time period, the reverberation power envelope substantially coincides across positions.

Owing to the reciprocity principle of acoustics, the same impulse responses and reverberation energy decay envelope would be obtained if source and receiver positions were swapped. By inspection of Fig. \ref{fig:pressureMap}, it is justified to deduct that the same decay envelope would be obtained regardless of source and receiver positions. Therefore, the \textit{Reverberation Initial Power} (RIP), referenced by extrapolation back to the time origin (time of emission), is independent of position. In \cite{jot1997analysis}, it is shown that the RIP is proportional to the inverse of the room's cubic volume and independent of frequency -- which, in turn, justifies the Reverb energy-vs-distance calculation illustrated in Fig. \ref{fig:reverbEnergy}.
 


\end{document}